\pdfoutput=1
\newcommand*{\ATLASLATEXPATH}{latex/}
\documentclass[cernpreprint,texlive=2016,txfonts,UKenglish]{latex/atlasdoc}
\usepackage{latex/atlaspackage}
\usepackage{latex/atlasbiblatex}

\usepackage{latex/atlascontribute}

\usepackage{latex/atlasphysics}
\graphicspath{{logos/}{figures/}}

\newcommand{\PbPb}{\mbox{Pb+Pb}}
\newcommand{\dAu}{\mbox{d+Au}}
\newcommand{\pA}{\mbox{$p$+A}}

\newcommand{\Njet}{\mbox{$N_{\mathrm{jet}}$}}
\newcommand{\Nch}{\mbox{$N_{\mathrm{ch}}$}}

\newcommand{\sqrtsnn}{\mbox{$\sqrt{s_{_\text{NN}}}$}}
\newcommand{\sqrts}{\mbox{$\sqrt{s}$}}

\newcommand{\pythia}{\textsc{Pythia}}
\newcommand{\pythiasix}{\textsc{Pythia}6}
\newcommand{\pythiaeight}{\textsc{Pythia}8}
\newcommand{\herwig}{\textsc{Herwig}++}

\newcommand{\RFour}{\mbox{$R = 0.4$}}

\newcommand{\RTwo}{\mbox{$R= 0.2$}}

\newcommand{\Dz}{\mbox{$D(z)$}}

\newcommand{\Dpt}{\mbox{$D(\pT)$}}

\newcommand{\pTtrue}{\mbox{$p_{\mathrm{T}}^{\mathrm{truth}}$}}

\newcommand{\Rdz}{\mbox{$R_{D(z)}$}}
\newcommand{\Rdpt}{\mbox{$R_{D(\pT)}$}}

\newcommand{\diff}{\mathrm{d}}

\newcommand{\pPb}{\mbox{$p$+Pb}}
\newcommand{\ptjet}{\mbox{$p_{\mathrm{T}}^{\mathrm{jet}}$}}

\newcommand{\etatrk}{\mbox{$\eta$}}

\newcommand{\ystar}{\mbox{$y^{*}_{\mathrm{jet}}$}}

\renewcommand{\_}{{\tt \char`\_}}  % works properly only in \tt mode (!)
\newcommand{\z}{\mbox{$z$}}

\AtlasTitle{Measurement of jet fragmentation in 5.02~\TeV\ proton--lead and proton--proton 
collisions with the ATLAS detector}

\author{The ATLAS Collaboration}

\AtlasRefCode{HION-2015-04}

\PreprintIdNumber{CERN-EP-2017-065}

\AtlasJournalRef{Nucl. Phys. A 978 (2018) 65}
\AtlasDOI{10.1016/j.nuclphysa.2018.07.006}

\AtlasAbstract{
 A measurement of the fragmentation functions of jets into 
 charged particles in \pPb\ collisions  
 and \pp\ collisions is presented.
 The analysis utilizes 28~nb$^{-1}$ of \pPb\ data and 26~pb$^{-1}$ of \pp\ data,
 both at \sqrtsnn~=~5.02~\TeV, collected in 2013 and
 2015, respectively, with the \mbox{ATLAS} detector at the LHC. 
 The measurement is reported in the centre-of-mass frame of
 the nucleon--nucleon system for jets
 in the rapidity range $|y^{*}|<$1.6 
 and with transverse momentum  45~$ < \pt <$~260~GeV. 
 Results are presented both as a function of the charged-particle 
 transverse momentum and as a function of the longitudinal momentum fraction of the 
 particle with respect to the jet.
 The \pp\ fragmentation functions are compared with results from Monte Carlo event generators and two theoretical models.
 The ratios of the \pPb\ to \pp\ fragmentation functions are found 
 to be consistent with unity.
}

\hypersetup{pdftitle={ATLAS document},pdfauthor={The ATLAS Collaboration}}

\begin{document}

\maketitle

%\tableofcontents

% List of contributors - print here or after the Bibliography.
%\PrintAtlasContribute{0.30}
%\clearpage

%%%%%%%%%%%%%%%%%%%%%%%%%%%%
\section{Introduction}
\label{sec:intro}

Heavy-ion collisions at the Large Hadron Collider (LHC) are performed in order to produce
and study the quark--gluon plasma (QGP), a phase of strongly interacting matter which emerges at very high
energy densities; a recent review can be found in Ref.~\cite{Roland:2014jsa}. Measurements of jets and jet properties in heavy-ion collisions are sensitive to the properties of the QGP.
In order to quantify jet modifications in heavy-ion collisions, proton--proton (\pp) collisions are often used
as a reference system.
Using this reference, rates of jet production in \PbPb\ collisions are observed to be reduced compared to that expected from the rates in \pp\ collisions, appropriately scaled to account for the nuclear thickness in \PbPb\ collisions~\cite{Aad:2014bxa,Khachatryan:2016jfl}. 
Charged-particle fragmentation functions are also observed to be modified in \PbPb\ collisions compared
to \pp\ collisions~\cite{Aad:2014wha,Aaboud:2017bzv,Chatrchyan:2014ava}. Both of these effects are interpreted as arising predominantly from the modification of the parton showering process in the final stages of the collision.

In addition to final-state differences emerging from the presence of the hot and dense matter, jet production in \PbPb\ collisions may also differ from that in \pp\ collisions due to effects arising from the presence of the large nucleus. For example, nucleons bound in a nucleus are expected to have a modified structure compared to the free nucleon~\cite{Salgado:2011wc}, and partons may lose energy in the nuclear environment before scattering~\cite{Kang:2015mta}. 
Proton--nucleus collisions are used to differentiate between initial- and final-state effects in \PbPb\ collisions. 
The inclusive jet production rate in proton--lead (\pPb) collisions at 5.02~\TeV\ 
was measured~\cite{ATLAS:2014cpa,Adam:2015hoa,Khachatryan:2016xdg} at the LHC 
and found to be only slightly modified after normalization by the nuclear thickness function.
Measurements made at the Relativistic Heavy Ion Collider with deuteron--gold collisions yield 
similar results~\cite{Adare:2015gla} (interestingly, Refs.~\cite{ATLAS:2014cpa,Adare:2015gla} observe
a centrality dependence to inclusive jet production).
High transverse momentum (\pT) charged hadrons originate from the fragmentation of jets and provide a complementary observable to that of jet production. The CMS Collaboration observed a small excess in the charged-particle spectrum measured in \pPb\ collisions for $\pT >$~20~\GeV\ particles compared to that expected from \pp\ collisions~\cite{Khachatryan:2016odn}.
Measurements of charged-particle fragmentation functions for jets in different \pT\ intervals in \pPb\ and \pp\ collisions are crucial for connecting the jet and charged-particle results.
Therefore, the measurements reported here are necessary both to establish a reference for jet fragmentation measurements in \PbPb\ collisions and to determine any modifications to jet fragmentation in \pPb\ collisions due to the presence of a large nucleus.

In recent years many of the features of \PbPb\ collisions which were interpreted as final state effects
due to hot nuclear matter were also observed in \pPb\ collisions at the LHC and in \dAu\ collisions at
RHIC.  These features include long-range hadron 
correlations~\cite{CMS:2012qk,Abelev:2012ola,Aad:2012gla,Adare:2013piz} and
a centrality-dependent reduction in the quarkonia 
yields~\cite{Adare:2013ezl,Adam:2016ohd,Sirunyan:2017mzd,Aaboud:2017cif}.
There is considerable debate about whether these features arise from the same source as in \PbPb\
collisions~\cite{Bozek:2013uha} or from other effects such as initial state gluon 
saturation~\cite{Dusling:2013qoz}.  
Measurements of jets in \pPb\ collisions showed no effects that would be attributable to hot nuclear matter, however additional measurements of jet properties in these collisions could help to constrain the source of the modifications observed in other observables.

In this paper, the jet momentum structure in \pp\ and \pPb\ collisions 
is studied using the distributions of charged particles associated with jets which have a transverse momentum \ptjet\ in the range 45 to 260~\GeV. 
Jets are reconstructed with the \antikt\ algorithm~\cite{Cacciari:2008qp}
using a radius parameter \RFour.
Charged particles are assigned to jets via an angular matching $\Delta R<0.4$,\footnote{
  ATLAS uses a  
right-handed coordinate system with its origin at the nominal
interaction point (IP) in the centre of the detector and the
$z$-axis along the beam pipe. The $x$-axis points from the IP to the
centre of the LHC ring, and the $y$-axis points upward. Cylindrical
coordinates $(r,\phi)$ are used in the transverse plane, $\phi$
being the azimuthal angle around the beam pipe. The pseudorapidity
is defined in terms of the polar angle $\theta$ as
$\eta=-\ln\tan(\theta/2)$. Rapidity is defined as $y = 0.5 \ln \frac{E+p_z}{E-p_z}$ where
$E$ and $p_z$ are the energy and the component of the momentum along the beam direction. Angular distance is measured in units of $\Delta R \equiv \sqrt{(\Delta\eta)^{2} + (\Delta\phi)^{2}}$.} where  $\Delta R$ is the angular distance between the jet axis and the charged-particle position. Results on the fragmentation functions are presented both as a function of
the ratio between the component of the particle transverse momentum parallel to the jet direction, and the jet \pT, $\z
 \equiv \pt \cos \Delta R \: / \:\ptjet$,\footnote{The $\Delta R$ is an approximation of the opening angle $\sqrt{(\Delta\theta)^{2} + (\Delta\phi)^{2}}$.} and as a function of the charged-particle transverse momentum with respect to the beam direction, \pt:
  \begin{equation}
\Dz \equiv \dfrac{1}{\Njet} \dfrac{\diff\Nch}{\diff z},
\label{eq:ffuncdef}
  \end{equation}
  and
  \begin{equation}
\Dpt \equiv \dfrac{1}{\Njet} \dfrac{\diff\Nch}{\diff \pt}.
\label{eq:ffuncdefpt}
  \end{equation}
The quantity \Nch\ is the number of charged particles and \Njet\ is the number of jets under consideration. The fragmentation functions are per-jet normalized.

The fragmentation functions are compared in \pPb\ and \pp\ collisions at a centre-of-mass energy of 5.02~\TeV.
In order to quantify any difference between \pPb\ and \pp\ collisions, the ratios of the fragmentation functions are measured:
\begin{equation}
\Rdz\ \equiv \frac{\Dz_{p\mathrm{Pb}}}{\Dz_{pp}}.
\label{eq:rdz} 
\end{equation}
In \PbPb\ collisions, such measurements are also presented as a function of charged-particle \pT~\cite{Aad:2014wha,Chatrchyan:2014ava} to explore the absolute \pT\ scale of the modifications and to reduce jet-related uncertainties. Thus, in addition to the more commonly used fragmentation functions as a function of $z$, this Letter also presents the analogous distributions and their ratios as a function of charged particle \pT:
\begin{equation}
\Rdpt \equiv \ \frac{\Dpt_{p\mathrm{Pb}}}{\Dpt_{pp}}.
\label{eq:rdpt}
\end{equation}

%\clearpage
%%%%%%%%%%%%%%%%%%%%%%%%%%%
\section{Experimental set-up}
\label{sec:setup}
The measurements presented here are performed using the ATLAS
calorimeter, inner detector,  trigger, and data acquisition systems
\cite{Aad:2008zzm}. The calorimeter system consists of a sampling liquid
argon (LAr) electromagnetic (EM) calorimeter covering $|\eta|<3.2$, a
steel--scintillator sampling hadronic calorimeter covering $|\eta| <1.7$, 
a LAr hadronic calorimeter covering $1.5 < |\eta| < 3.2$, and
two LAr forward calorimeters (FCal) covering $3.2 < |\eta| <
4.9$. The hadronic calorimeter has three
sampling layers longitudinal in shower depth.
The EM calorimeters are segmented longitudinally in shower depth into three
layers plus an additional pre-sampler layer. The EM calorimeter 
has a granularity that varies with layer and pseudorapidity, but which
is generally much finer than that of the hadronic
calorimeter. The minimum-bias trigger scintillators (MBTS)~\cite{Aad:2008zzm} detect charged particles over $2.1 < |\eta| < 3.9$ using two segmented counters placed at $z = \pm 3.6$~m. Each counter provides measurements of both the pulse heights and the arrival times of ionization energy deposits. 

A two-level trigger system was used to select the \pPb\ and \pp\ collisions analysed here. 
The first, the hardware-based trigger stage Level-1, is implemented with custom electronics. 
The second level is the software-based High Level Trigger (HLT). 
Jet events were selected by the HLT with Level-1 seeds from jet, minimum-bias, and 
total-energy triggers. The total-energy trigger required a total transverse energy measured in the calorimeter of greater than 5~\GeV.
The HLT jet trigger operated a jet reconstruction 
algorithm similar to that applied in the offline analysis and  selected events containing jets with 
transverse energy thresholds ranging from~20~\GeV\ to~75~\GeV\ in \pPb\ collisions and up to 85~\GeV\ in \pp\ collisions.
In both the \pp\ and \pPb\ collisions, the highest-threshold jet trigger sampled the full 
delivered luminosity.
Minimum-bias \pPb\ events were required to have at least one hit in a counter on each side of the MBTS detector at the Level-1 trigger.

The inner detector measures charged-particle tracks within the pseudorapidity interval 
$|\eta|<2.5$ using a combination of silicon pixel detectors, silicon
microstrip detectors (SCT), and a straw-tube transition radiation
tracker (TRT),
all immersed in a 2~T axial magnetic field~\cite{Aad:2008zzm}. Each of the three detectors is composed of a barrel and two symmetric end-cap sections.
The pixel detector 
is composed of three layers of sensors with a nominal
pixel size of $50~{\mathrm \mu m} \, \times \, 400~{\mathrm \mu m}$. 
Following the \pPb\ data-taking and prior to the  5~\TeV\ \pp\ data-taking
 an additional silicon tracking layer, the ``insertable B-layer'' (IBL)~\cite{ibl1}, was
installed closer to the interaction point than the other three layers.
 The 
SCT barrel section contains four layers of modules with 80~$\mu{\mathrm{m}}$ pitch sensors on both sides, and each end-cap consists of nine
layers of double-sided modules with radial strips having a mean pitch
of 80~$\mu\mathrm{m}$. The two sides of each SCT layer in both the barrel
and the end-caps have a relative stereo angle of 40~mrad. The TRT contains up to 73 (160) layers of
staggered straws interleaved with fibres in the barrel (end-cap).

%%%%%%%%%%%%%%%%%%%%%%%%%%%
\section{Event selection and data sets}
\label{sec:data}
The \pPb\ data used in this analysis were recorded in 2013. The LHC was configured
with a 4 \TeV\ proton beam and a 1.57~\TeV\ per nucleon Pb beam  producing collisions
with \sqrtsnn~=~5.02~\TeV\ and a rapidity shift of the centre-of-mass frame, $\Delta y=0.465$, 
relative to the laboratory frame. 
The data collection 
was split into two periods with opposite beam configurations. The first period
consists of approximately 55\% of the integrated luminosity with the Pb beam travelling toward positive rapidity and the proton beam to negative rapidity. The remaining data were taken with the beams of protons and Pb nuclei swapped. 
The total \pPb\ integrated luminosity 
is 28~nb$^{-1}$. Approximately 26~pb$^{-1}$ of \sqrts~=~5.02~\TeV\ \pp\ data from 2015 
was used. The instantaneous luminosity conditions provided 
by the LHC resulted in an average number of \pPb\ interactions per bunch crossing 
of 0.03. During \pp\ data-taking, the average number of interactions per
bunch crossing  varied from 0.6 to 1.3.

The \pPb\ events selected are required to have a reconstructed vertex, at least one hit in each 
MBTS detector, and a time difference measured between the two MBTS sides of less than 10~ns. The \pp\ events used in this analysis are required to have a reconstructed vertex; no requirement on the signal 
in the MBTS detector is imposed.  
In \pPb\ collisions the event centrality is determined by the FCal in the Pb-going direction as
in Ref.~\cite{ATLAS:2014cpa}.  The \pPb\ events used here belong to the 0--90\% centrality interval.

The performance of the ATLAS detector and offline analysis in measuring fragmentation functions in \pPb\  collisions is evaluated using a sample of Monte Carlo (MC) events obtained by overlaying 
simulated hard-scattering \pp\ events generated with 
\pythia\ version 6.423 (\pythiasix)~\cite{Sjostrand:2006za} onto minimum-bias \pPb\ events recorded during the same data-taking period. 
A sample consisting of 2.4$\times 10^{7}$  \pp\ events 
is generated with \pythiasix\ using 
parameter values from the AUET2B tune~\cite{ATLAS-PHYS-PUB-2011-009} and the CTEQ6L1 parton distribution
function (PDF) set~\cite{Pumplin:2002vw}, at $\sqrts=5.02$~\TeV\ and  
with a rapidity shift equivalent to that in the \pPb\ collisions is used in the overlay procedure.
About half of the events are simulated with one beam configuration and the second half with the other.
 The detector response 
is simulated using GEANT4~\cite{Agostinelli:2002hh,Aad:2010ah}, and the simulated hits are combined 
with those from the data event.
An additional sample of 2.6$\times10^{7}$ \pp\ hard-scattering events simulated with \pythia\ version 8.212 (\pythiaeight)~\cite{Sjostrand:2014zea} at $\sqrts=5.02$~\TeV\ with the A14 tune~\cite{ATLAS2014021} and NNPDF23LO PDF set~\cite{Ball:2012cx} is used to evaluate the performance for measuring fragmentation functions in the 2015 \pp\ data.
Finally, fragmentation functions at generator-level evaluated from 
1.5$\times 10^{7}$~5.02~\TeV\ \pp\ events~\cite{Bahr:2008pv}  
generated with \herwig\ using the UEEE5 tune~\cite{Gieseke:2012ft} and the CTEQ6L1 PDFs~\cite{Pumplin:2002vw}
are compared to the fragmentation function measured in 5.02~\TeV\ \pp\ data.

%%%%%%%%%%%%%%%%%%%%%%%%%%%
\section{Jet and track selection}
\label{sec:reconstruction}
Jets are reconstructed with the same heavy-ion jet reconstruction algorithm used in previous measurements in \pPb\ collisions~\cite{ATLAS:2014cpa}. The \antikt\ algorithm~\cite{Cacciari:2008qp} is first run in four-momentum recombination mode using as input the signal in $\Delta \eta \times \Delta \phi = 0.1\times 0.1$  calorimeter towers with the \antikt\ radius parameter $R$ set to 0.4 and 0.2
	($R=$~0.4 jets are used for the main analysis and the $R=$~0.2 jets are used to improve the jet position resolution
	as discussed below). The energies in the towers are obtained by summing the
	energies of calorimeter cells at the electromagnetic energy scale within the tower boundaries. Then,
	  an iterative procedure is used to estimate the layer- and $\eta$-dependent
	underlying event (UE) transverse energy density, while excluding the regions populated by jets.
	  The UE transverse energy is subtracted from each calorimeter cell and the four-momentum of the jet is updated accordingly.
	Then, a jet $\eta$- and \pT-dependent  correction factor derived from the simulation samples is applied to correct the jet momentum for the calorimeter response. Additionally, the jet energies were corrected by a multiplicative factor derived in \textit{in situ} studies of the transverse momentum balance of 
	jets recoiling against photons, $Z$ bosons, and jets in other regions of the calorimeter~\cite{Aad:2011he,Run2jetpubnote}. This \textit{in situ} calibration, which typically differed from unity by a few percent, accounts for differences between the simulated detector response and data.

Jets are required to have jet centre-of-mass rapidity, $|\ystar|<1.6$,\footnote{The jet centre-of-mass rapidity $\ystar$ is defined as $\ystar \equiv y_{\mathrm{jet}} - \Delta y$ where $y_{\mathrm{jet}}$ is the jet rapidity in the ATLAS rest frame and $\Delta y$ is the rapidity shift of the centre-of-mass frame.} 
which is the largest symmetric overlap between the two collision systems for which there is full charged-particle tracking coverage
within a jet cone of size $R=$~0.4.
To prevent neighbouring jets from distorting the measurement of the fragmentation functions, 
jets are rejected if there is another jet with higher \pt within a distance $\delta R = 1.0$,
where $\delta R$ is the distance between the two jet axes. 
To reduce the effects of the broadening of the jet position
measurement due to the UE, for \RFour\ jets, the jet direction is taken from that of the closest
matching \RTwo\ jet within $\delta R =0.3$ when such a matching jet is found
(this procedure has been previously used in Ref.~\cite{Aaboud:2017bzv}).
All jets included in the analysis are required to have \pt\ sufficiently large for the jet 
trigger efficiency to be higher than 99\%.
Reconstructed jets which consist only of isolated high-\pT\ electrons~\cite{Aad:2014fxa} from electroweak bosons are excluded from this analysis. 

The MC samples are used to evaluate the jet reconstruction performance and to correct the measured
distributions for detector effects.  
The \pPb\ jet reconstruction performance is described in Ref.~\cite{ATLAS:2014cpa}; the jet reconstruction performance in \pp\ collisions is found to be similar to that in \pPb\ collisions.
In the MC samples, the kinematics of the particle-level jets are reconstructed from primary particles\footnote{Primary particles are defined as particles with a mean lifetime $\tau>0.3\times 10^{-10}$~s either directly produced in \pp\ interactions or from subsequent decays of particles with a shorter lifetime. All other particles are considered to be secondary.} with the \antikt\ algorithm with radius parameter
\RFour. In these studies, particle-level jets are matched to reconstructed jets with a $\Delta R <$~0.2.

Tracks used in the analysis of \pPb\ collisions are required 
to have at least one hit in the pixel detector and at least six hits in the SCT. 
Tracks used in the analysis of \pp\ collisions are required to have at least 9 or 11 total silicon hits for 
|\etatrk| < 1.65 or |\etatrk| > 1.65, respectively, 
including both the pixel layers and the SCT.  This includes a hit
in the first (first or second) pixel layer if expected from the track trajectory for the \pPb\ (\pp) data.
All tracks used in this analysis are required to have $\pt > 1$~\GeV.
In order to suppress the contribution of secondary particles, the distance of closest approach of the track to the primary vertex is required to be less than 1.5~mm along the beam axis and less than a value which varies from approximately 0.6~mm at \pt~=~1~\GeV\ to approximately 0.2~mm at \pt~=~20~\GeV\ in the transverse plane.

\begin{figure}[ht]
\centerline{
\begin{tabular}{cc}
\includegraphics[width=0.45\textwidth]{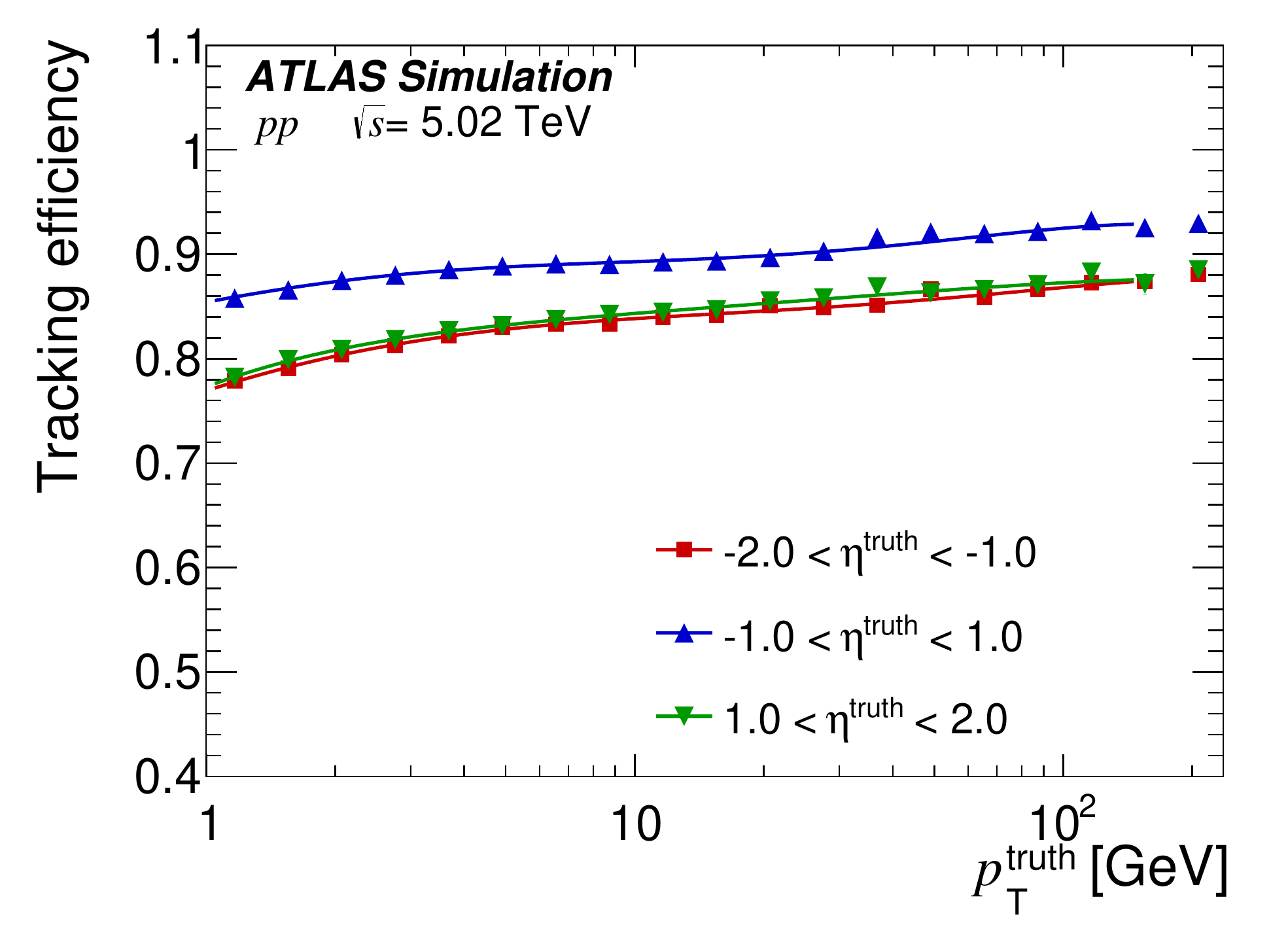} & 
\includegraphics[width=0.45\textwidth]{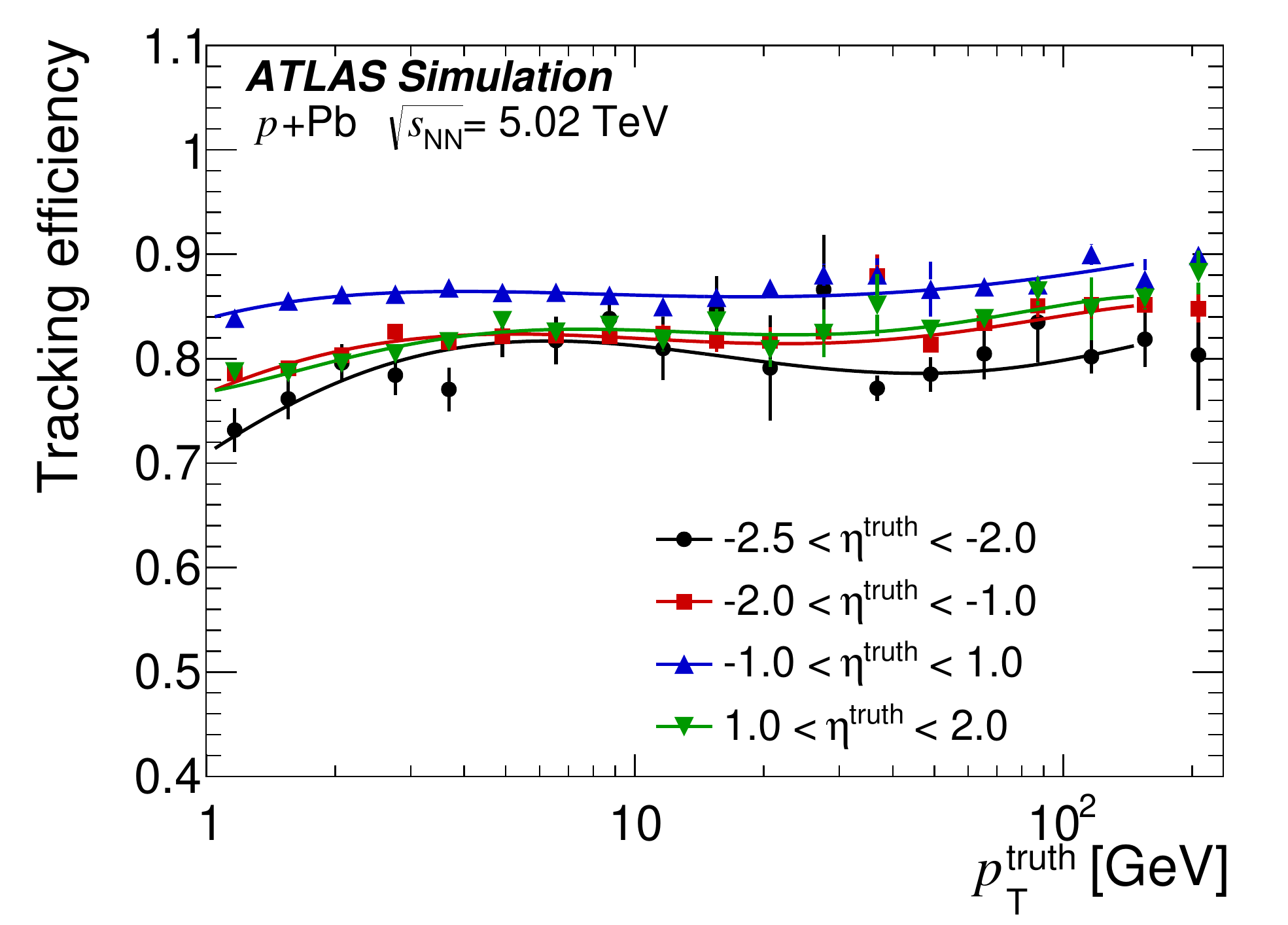} \\
\end{tabular} }
\caption{  Tracking efficiency as a function the primary particle momentum at generation level, \pTtrue,  in \pp\ collisions (left) and in \pPb\ collisions for one of the two beam configurations (right). The different sets of points show the primary particle pseudorapidity, $\eta^{\mathrm{truth}}$, intervals in which  the track reconstruction efficiency has been performed. The different $\eta^{\mathrm{truth}}$ intervals in \pp\ and \pPb\ plots reflect the different regions of the tracking system used in the two cases due to the boosted \pPb\ system. The solid curves show parameterizations of efficiencies.
}
\label{fig:trackingeff}
\end{figure}

The efficiency for reconstructing charged particles within jets in \pPb\ and \pp\ collisions
is evaluated using \pythiasix\ and \pythiaeight\ MC samples, respectively,
and is computed by matching the reconstructed tracks to generator-level primary particles. The association is done based on contributions of generator-level particles to the hits in the detector layers. A reconstructed track is matched to a generator-level particle if it contains hits produced primarily by this particle~\cite{Aad:2010ah}. 
The efficiencies are determined separately for the two  \pPb\ running configurations because
the $\eta$ regions of the detector used for the track measurement are different for the two
beam configurations.
The charged-particle reconstruction efficiencies 
as a function of the primary particle's transverse momentum, 
\pTtrue, in coarse $\eta^{\mathrm{truth}}$ intervals, are shown in Figure~\ref{fig:trackingeff} in \pp\ and \pPb\ collisions. The \pTtrue\ dependence of the efficiencies is parameterized using a 
fifth-order polynomial in $\log(\pTtrue)$ which
describes the efficiency behaviour in the range of particle \pTtrue\ from 1.0 to 150~\GeV. The tracking efficiency is observed to be constant above 150 GeV and a constant efficiency value is used for particles with $\pTtrue > 150$~\GeV\ due to the limited size of the MC samples in that phase space region. 
To account for finer scale variations of the tracking 
efficiency with pseudorapidity, the parameterizations are multiplied by an $\eta$-dependent scale factor evaluated in $\eta^{\mathrm{truth}}$ intervals of 0.1 units in coarse \pTtrue\ intervals. 
The dependence of the charged-particle efficiency on \ptjet\ is found to be negligible for the \ptjet\ selections used here.  The measured \Dz\ and \Dpt\ distributions are corrected for the contribution of reconstructed ``fake'' tracks which cannot be matched to a generated primary particle in the MC samples produced without minimum bias interactions overlaid and the residual contribution of tracks matched to secondary particles. %are together called the contribution from ``fake'' tracks.
The fraction of secondary and fake tracks is found to be below 2\% of the tracks that pass the selection in any track and jet kinematic region. The contribution from these tracks to the fragmentation functions is subtracted from the measured fragmentation functions in both the \pp\ and \pPb\ collisions.

%%%%%%%%%%%%%%%%%%%%%%%%%%%
\section{Analysis procedure}
\label{sec:analysis}
Reconstructed charged particle tracks are associated with a reconstructed jet if they 
fall within $\Delta R = 0.4$ of the jet axis.  For each of
these particles the momentum fraction, \z, is calculated.
The measured fragmentation functions are constructed as:
\begin{equation}
   \Dz_{\mathrm{meas}} \equiv \frac{1}{\Njet} \frac{1}{\varepsilon(\eta,\pt)} \frac{\Delta \Nch(\z)}{\Delta\z}
\label{eq:dzmeas}
\end{equation}
and
\begin{equation}
   \Dpt_{\mathrm{meas}} \equiv \frac{1}{\Njet} \frac{1}{\varepsilon(\eta,\pt)} \frac{\Delta \Nch(\pt)}{\Delta\pt},
\label{eq:dptmeas}
\end{equation}

where $\varepsilon(\eta,\pt)$ is the track reconstruction efficiency, and \Njet\ is the 
total number of jets in a given \ptjet\ bin.  The quantities $\Delta \Nch(\z)$ and $\Delta \Nch(\pt)$ are
the numbers of associated tracks within the given \z\ or \pt\ range, respectively. The efficiency
correction is applied on a track-by-track basis, assuming $\pt = \pTtrue$. While that assumption is not strictly valid, the efficiency varies sufficiently slowly with $\pTtrue$ that the error introduced by this assumption is negligible.

In \pPb\ collisions the UE contribution to the fragmentation functions from charged particles not associated with the jet constitutes a background that needs to be subtracted. It originates in soft interactions that accompany the hard process in the same \pPb\ collision and depends on charged-particle \pT\ and $\eta$. This background is determined event by event for each measured jet by using a grid of $\Delta R = 0.4$ cones that span the full coverage of the inner detector. The cones have a fixed distance between their centres chosen such that the coverage of the inner detector is maximized while the cones do not overlap each other. Any such cone containing a charged particle with $\pT>3.5$~\GeV\ is assumed to be associated with a real jet and is excluded from the UE contribution. The 3.5~\GeV\ threshold is derived from studies of UE contribution in MC samples. The estimated contribution from UE particles in each cone is corrected to account for differences in the average UE particle yield at a given \pt\ between the $\eta$ position of the cone and the $\eta$ position of the jet. The correction is based on a parameterization of the \pt\ and $\eta$ dependence of charged-particle yields in minimum-bias collisions. The resulting UE contribution is evaluated for charged particles in the transverse momentum interval of $1 < \pt\ < 3.5$~\GeV\ and averaged over all cones. The UE contribution is further corrected for the correlation between the actual UE yield within the jet cone and the jet energy resolution discussed in Ref.~\cite{Aaboud:2017bzv}. This effect is corrected by a multiplicative correction factor, dependent on the track \pt\ (or $z$) and the jet \pT. The correction is estimated in MC samples as the ratio of the UE contribution calculated from tracks within the area of a jet that do not have an associated generator-level particle to the UE contribution estimated by the cone method. Corrected UE contributions are then subtracted from measured distributions. The maximum size of the UE contribution is 20\% for the lowest track \pt\ (or $z$). The UE from the cone method is compared to an alternative UE estimation and the difference is found to be negligible. The subtracted UE contribution has no azimuthal variation in \pPb\ collisions
and no UE subtraction is performed for the \pp\ measurement due to negligible UE contribution (less
than 2\% over the entire kinematic range measured).

The measured \Dz\ and \Dpt\ distributions are corrected for detector effects by means of a two-dimensional 
Bayesian unfolding procedure~\cite{DAgostini:1994zf} using the RooUnfold package~\cite{Adye:2011gm}. The unfolding procedure removes the effect of bin migration due to the jet energy and the track momentum resolutions. Using the MC samples, four-dimensional response matrices are created using the particle-level and reconstructed \ptjet, and generator-level and reconstructed track \pt (\z). 
Separate unfolding matrices are constructed for the \pPb\ and \pp\ data.  An independent bin-by-bin unfolding procedure is used to correct the measured \ptjet\ spectra,
which is used to normalize the unfolded fragmentation functions by the number of jets.
The response matrices are reweighted such that the shapes of the measured fragmentation functions and jet spectra in the simulation match those in the data.
The number of iterations in the Bayesian unfolding is selected to be the minimum number for which
the relative change in the fragmentation function at
$z=0.1$ is smaller than 0.2\% per additional iteration in all \ptjet\ bins.
This condition ensures the stability of the unfolding and minimizes statistical fluctuations due to the unfolding
in the high \z\ and \pT\ regions.  The resulting number of iterations is driven by the low \ptjet\ intervals, which require the most iterations to converge.  The systematic uncertainty due to the unfolding is typically much larger than the impact of the stability requirement, especially for the lowest \ptjet\ values used in this analysis (discussed in 
Section~\ref{sec:systematics}).
Following this criterion, 14 iterations are used for both the \pPb\ and \pp\ data sets.
The analysis procedure is tested by dividing the MC event sample in half and using one
half to generate response matrices with which the other half is unfolded.  
Good recovery of the generator-level distributions is observed for the unfolded events and the deviations from perfect closure are incorporated into the systematic uncertainties.

%%%%%%%%%%%%%%%%%%%%%%%%%%%
\section{Systematic uncertainties}
\label{sec:systematics}
\begin{figure}
\centerline{
\begin{tabular}{cc}
\includegraphics[width=0.45\textwidth]{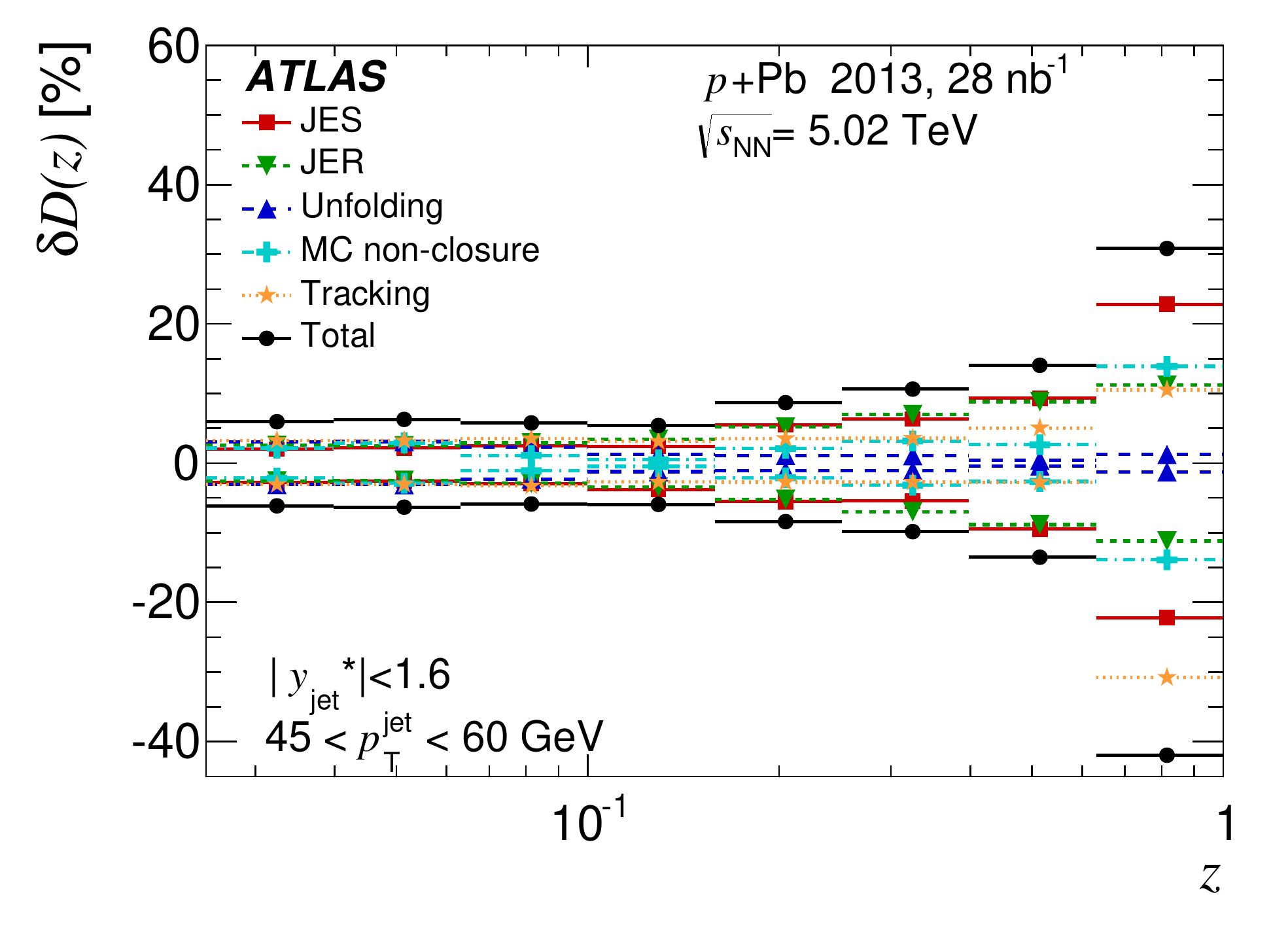} &
\includegraphics[width=0.45\textwidth]{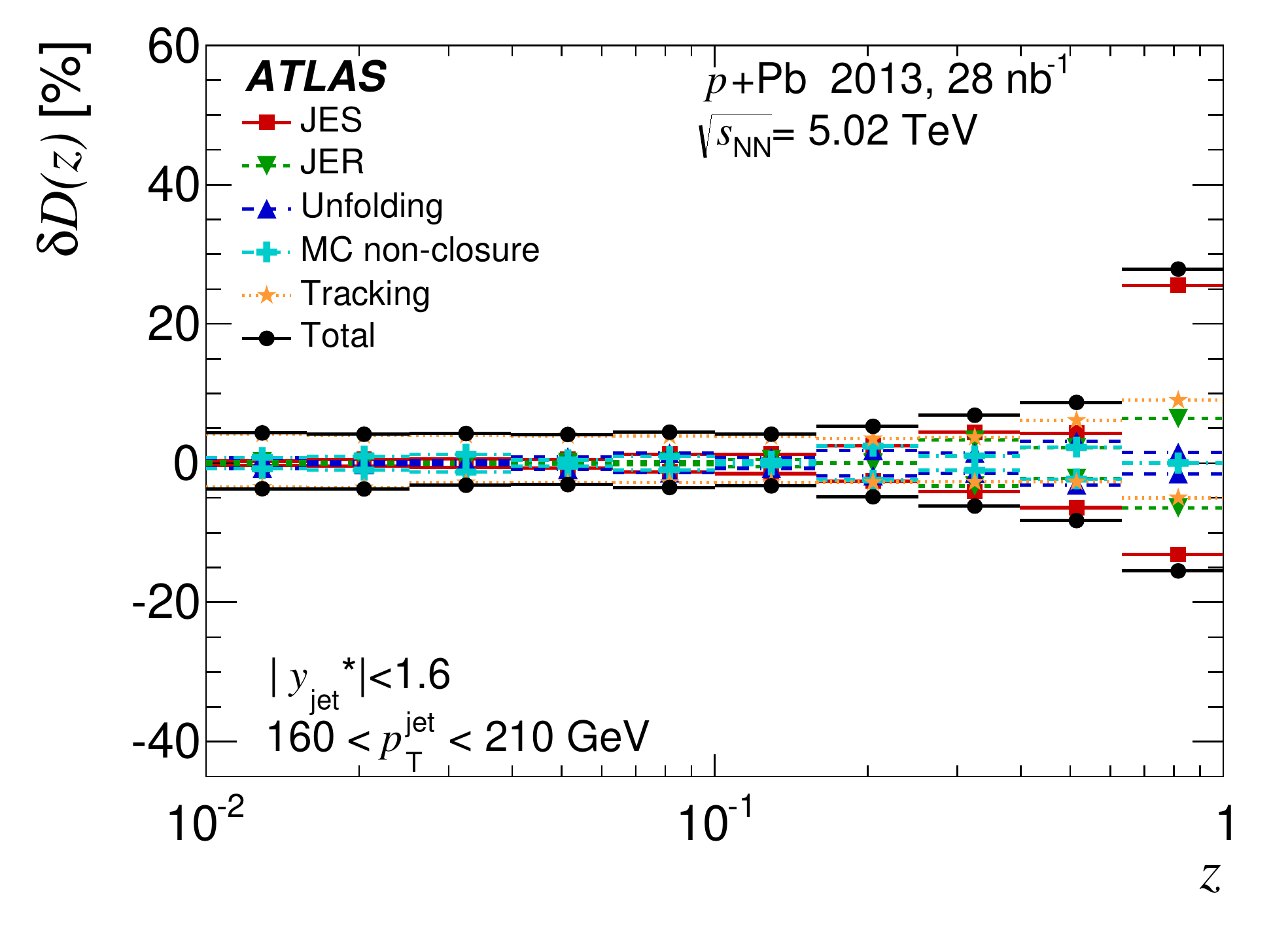} \\
\includegraphics[width=0.45\textwidth]{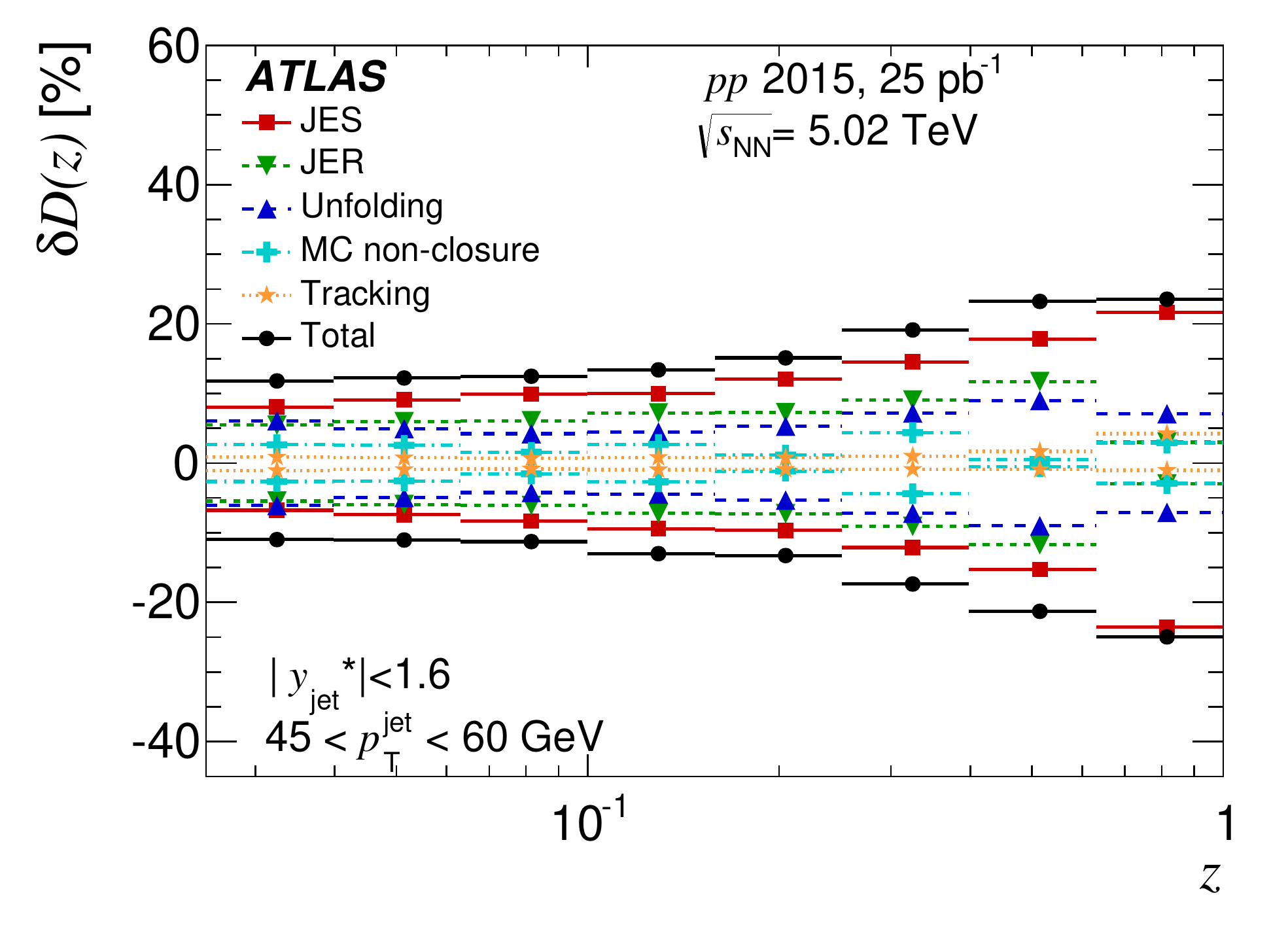} &
\includegraphics[width=0.45\textwidth]{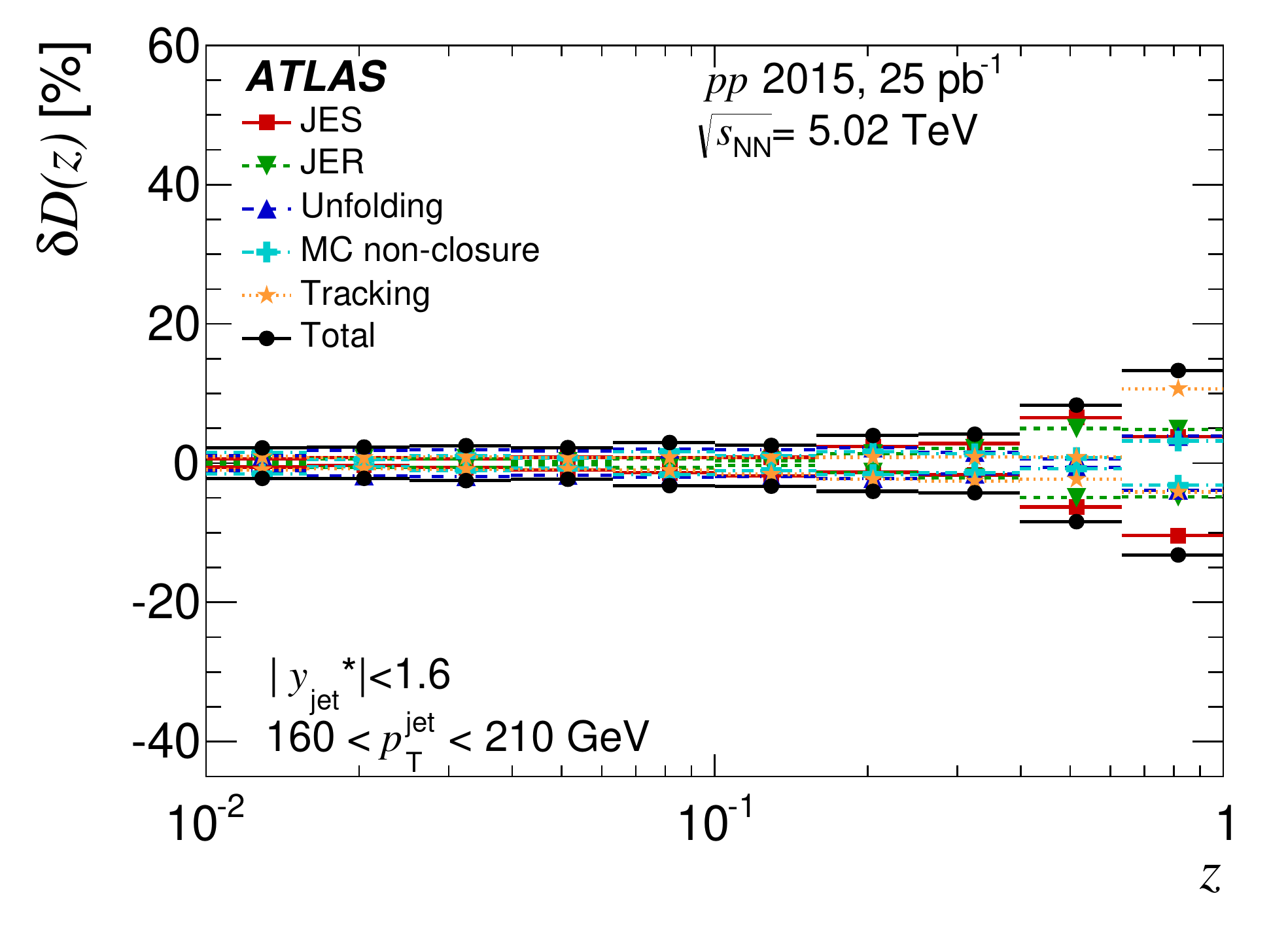} \\
\end{tabular}}
\caption{
Summary of the systematic uncertainties in the fragmentation function, $D(z)$, in \pPb\ collisions (top) and \pp\ collisions (bottom) for jets in the 45--60~\GeV\ \ptjet\ interval (left) and in the 160--210~\GeV\ \ptjet\ interval (right). The systematic uncertainties due to JES, JER, unfolding, MC 
non-closure and tracking are shown along with the total systematic uncertainty from all sources.   
}
\label{Fig:Systematics_1}
\end{figure}

\begin{figure}
\centerline{
\begin{tabular}{cc}
\includegraphics[width=0.45\textwidth]{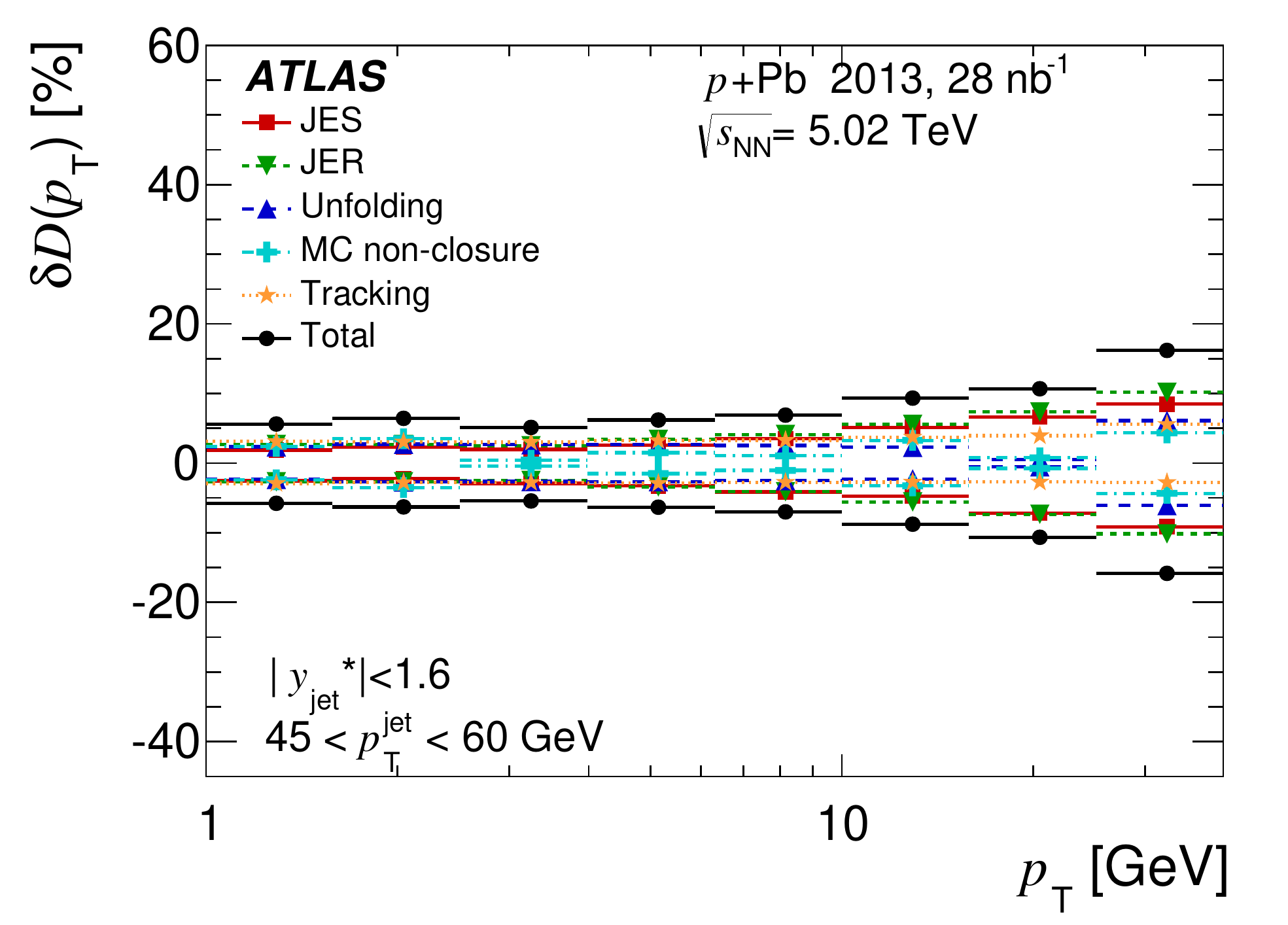} &
\includegraphics[width=0.45\textwidth]{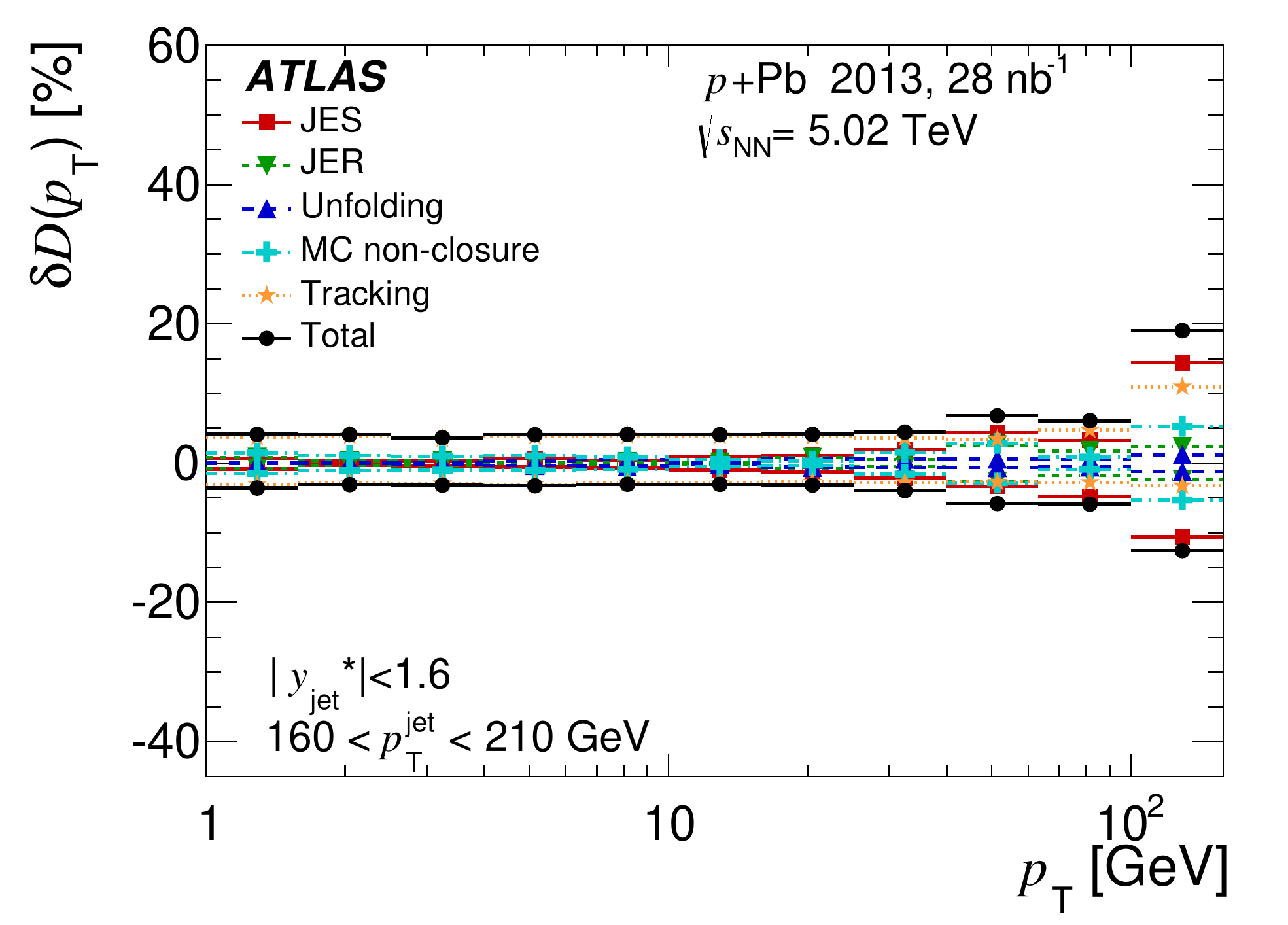} \\
\includegraphics[width=0.45\textwidth]{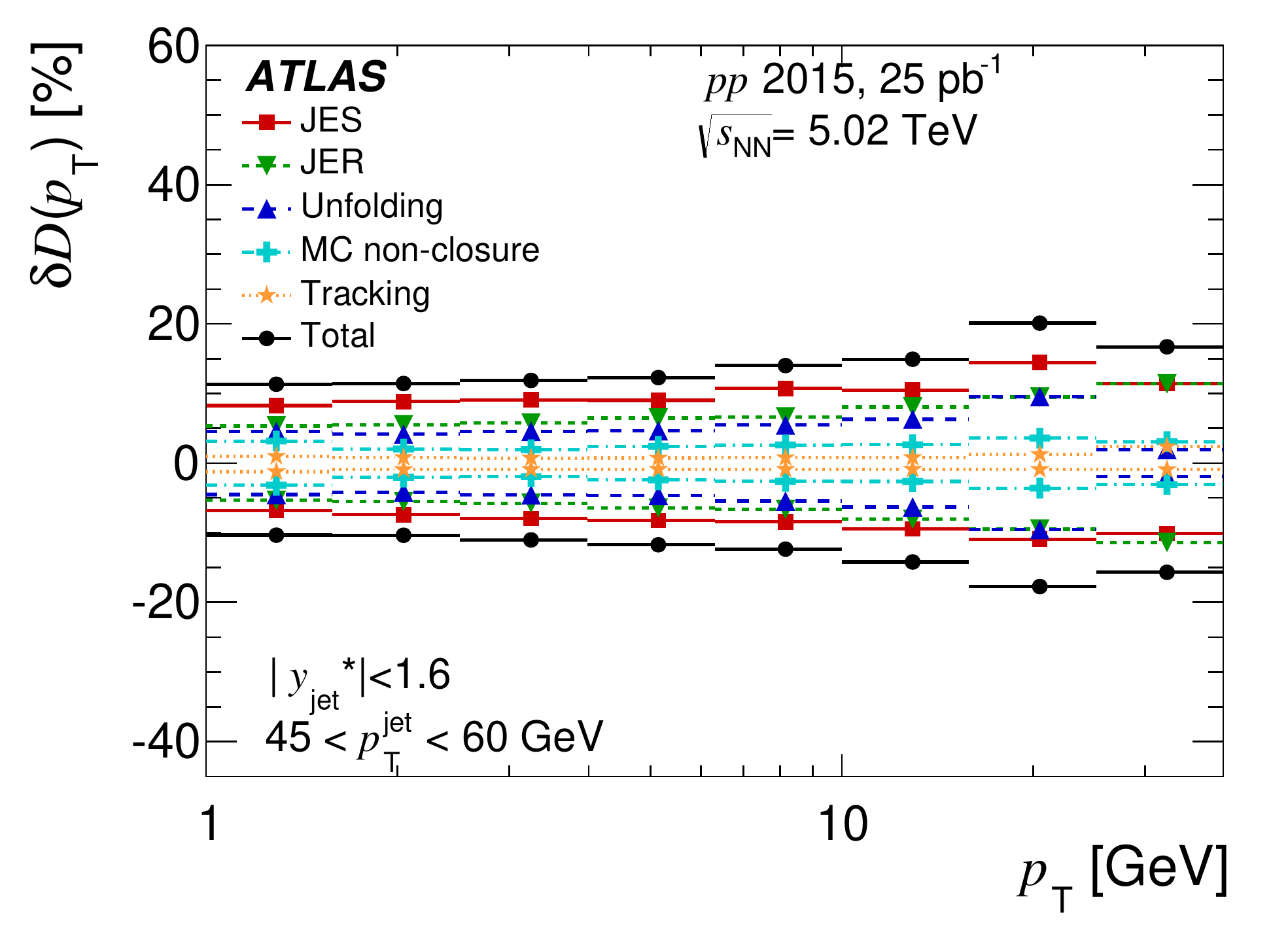} &
\includegraphics[width=0.45\textwidth]{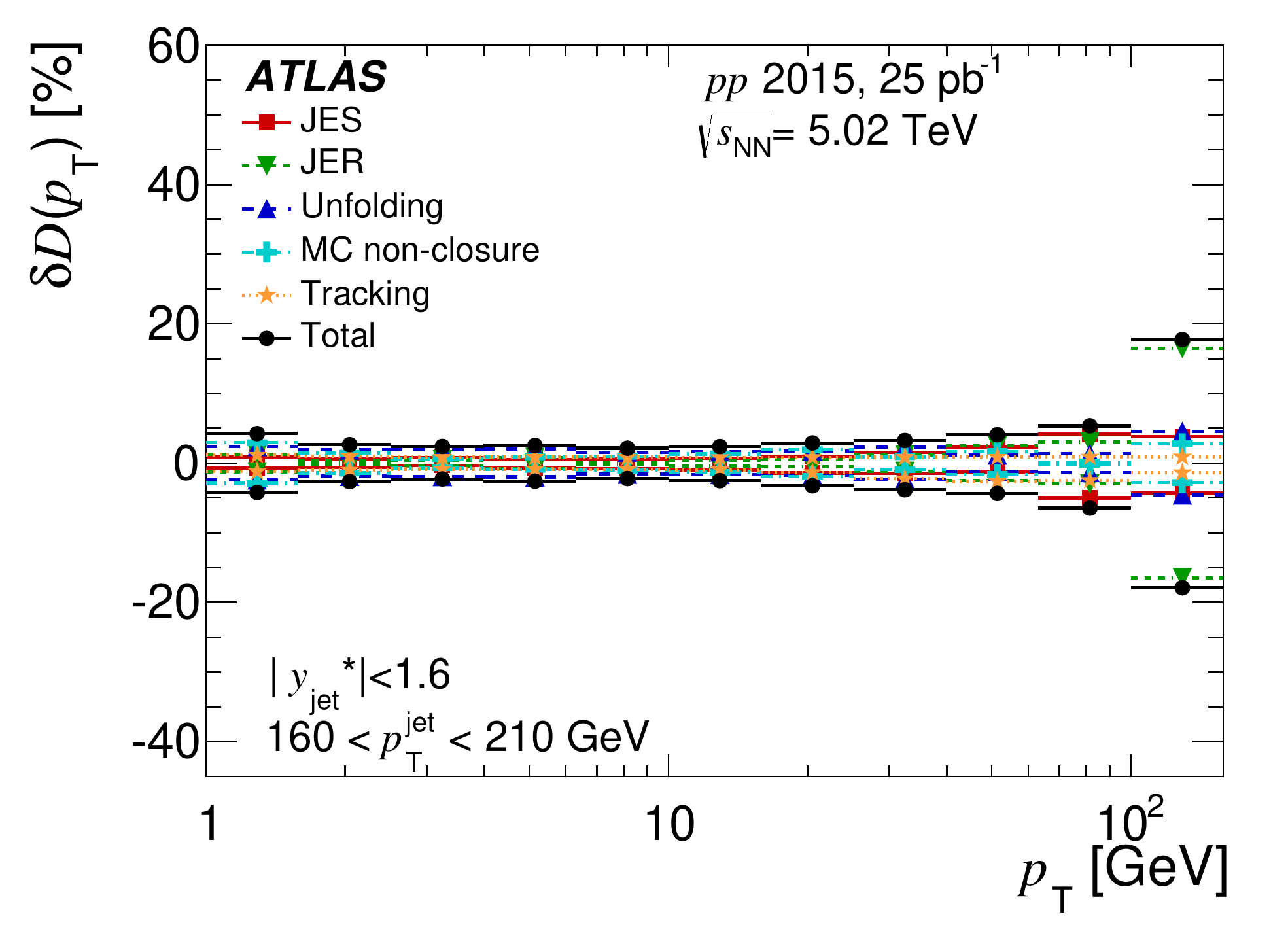} \\
\end{tabular}}
\caption{
Summary of the systematic uncertainties in the fragmentation function, $D(\pT)$, in \pPb\ collisions (top) and \pp\ collisions (bottom) for jets in the 45--60~\GeV\ \ptjet\ interval (left) and in the 160--210~\GeV\ \ptjet\ interval (right). The systematic uncertainties due to JES, JER, unfolding,  
MC non-closure and tracking are shown along with the total systematic uncertainty from all sources.
}
\label{Fig:Systematics_2}
\end{figure}

The systematic uncertainties in the measurement of the fragmentation functions and their ratios
are described in this section.  The following sources of systematic uncertainty in the measurement of the fragmentation functions and their ratios are considered:
the jet energy scale (JES), the jet energy resolution (JER), 
the dependence of the unfolded results on the choice of the starting MC distributions, 
the residual non-closure of the unfolding and the tracking-related uncertainties.
For each variation reflecting a systematic uncertainty the fragmentation functions are re-evaluated and the
difference between the varied and nominal fragmentation functions is used as
an estimate of the uncertainty.
The systematic uncertainties in the 
$D(z)$ and $D(\pT)$ measurements in both collision systems are summarized in 
Figures~\ref{Fig:Systematics_1}\ and~\ref{Fig:Systematics_2}, respectively, 
for two different jet \pt\ bins.  The systematic uncertainties from each source 
are taken as uncorrelated and combined in quadrature to obtain the total systematic uncertainty. 

The JES uncertainty is determined from \textit{in situ} studies of the calorimeter 
response~\cite{Aad:2011he,HIjesnote,Aaboud:2017jcu}, and studies of the relative energy-scale difference between the jet reconstruction procedure in heavy-ion collisions and the procedure used in \pp\ 
collisions~\cite{Aad:2014bia}. 
The impact of the JES uncertainty on the measured distributions is evaluated by constructing new response matrices where all reconstructed jet transverse momenta are shifted by $\pm 1$~standard deviation ($\pm 1 \sigma$) of the JES uncertainty. The data are then unfolded with these matrices. Each component that contributes to the JES uncertainty is varied separately. In total, 45 and 51 independent systematic components constitute the full JES uncertainty in the analysis of \pPb\ and \pp\ collisions, respectively. These components are uncorrelated among each other within the data set and fully correlated across \pT\ and $\eta$. The JES uncertainty increases with increasing $z$ and particle \pT\ at fixed \ptjet\ and decreases with increasing \ptjet. 

The uncertainty in the fragmentation functions due to the JER is estimated by repeating the unfolding procedure with modified response matrices, where the resolution of the reconstructed jet \ptjet\ is broadened by Gaussian smearing. The smearing factor is evaluated using an \textit{in situ} technique involving studies of dijet energy balance~~\cite{Aad:2012ag,JERConfNote}. The systematic uncertainty 
due to the JER 
increases with increasing $z$ and particle \pT\ at fixed \ptjet\ and decreases with increasing \ptjet.

The unfolding uncertainty is estimated by generating the response matrices from the MC distributions without reweighting to match the shapes of the reconstructed data in \ptjet\ and $D(\z)$ or $D(\pT)$.
Conservatively, an additional uncertainty to account for possible residual limitations in the analysis procedure was assigned by evaluating the non-closure of the unfolded distributions in simulations, as described in Section~\ref{sec:analysis}. The magnitude of both of these uncertainties is typically below 5\% except for the highest $z$ and track \pT\ bins.  

The uncertainties related to the track reconstruction and selection originate from several sources.
Uncertainties related to the rate of secondary and fake tracks, the material description in the simulation, and the track's transverse momentum were obtained from studies in data and simulation described in Ref.~\cite{ATL-PHYS-PUB-2015-051}.
The systematic uncertainty in the secondary-track and fake-track rate is 30\% in \pp\ collisions and
50\% in \pPb. The contamination by secondary and fake tracks is at most 2\%, the resulting uncertainty in the fragmentation functions is at most 1\%.
The sensitivity of the tracking efficiency to the description of the 
inactive material in the MC samples is evaluated by varying the material description.
This uncertainty is between
0.5 and 2\% (depending on track \etatrk) in the track \pT\ range used in the analysis. 
Uncertainty in the tracking efficiency due to the high local 
track density in the cores of jets is 0.4\%~\cite{ATL-PHYS-PUB-2016-007} for all \ptjet\ 
selections in this analysis.
The uncertainty due to the track selection criteria is evaluated by repeating 
the analysis with an additional requirement on the significance
of the distance of closest approach of the track to the primary vertex. This uncertainty affects both the track reconstruction efficiency and the rate of secondary and fake tracks. The resulting uncertainty typically varies from 1\% at low track \pT\ and low \z\ to 5\% at high track \pT\ and high \z.
The systematic uncertainties in the fragmentation functions due to the parameterization of the efficiency corrections is less than 1\%.
An additional uncertainty takes into account a possible residual misalignment of the tracking detectors in \pp\ data-taking. The alignment in this data was checked \textit{in situ} with $Z\rightarrow \mu^{+}\mu^{-}$ events, and thus a track-\pt\-dependent uncertainty arises from the finite size of this sample.  The resulting uncertainties in the fragmentation functions are typically smaller than 1\% except at large $z$ where they are as large as 4\%.
Finally, the track-to-particle matching requirements are varied. This variation affects the track reconstruction efficiency, the track momentum resolution, and the rate of secondary and fake tracks. The resulting uncertainties in the fragmentation functions are smaller than 1\%.
After deriving new response matrices and efficiency corrections, the resulting systematic uncertainty in the fragmentation functions is found to be less than 0.5\%.
The tracking uncertainties shown in Figures~\ref{Fig:Systematics_1} and~\ref{Fig:Systematics_2} include all the above explained track-related systematic uncertainties added in quadrature.

\begin{figure}
\centerline{
\begin{tabular}{cc}
\includegraphics[width=0.45\textwidth]{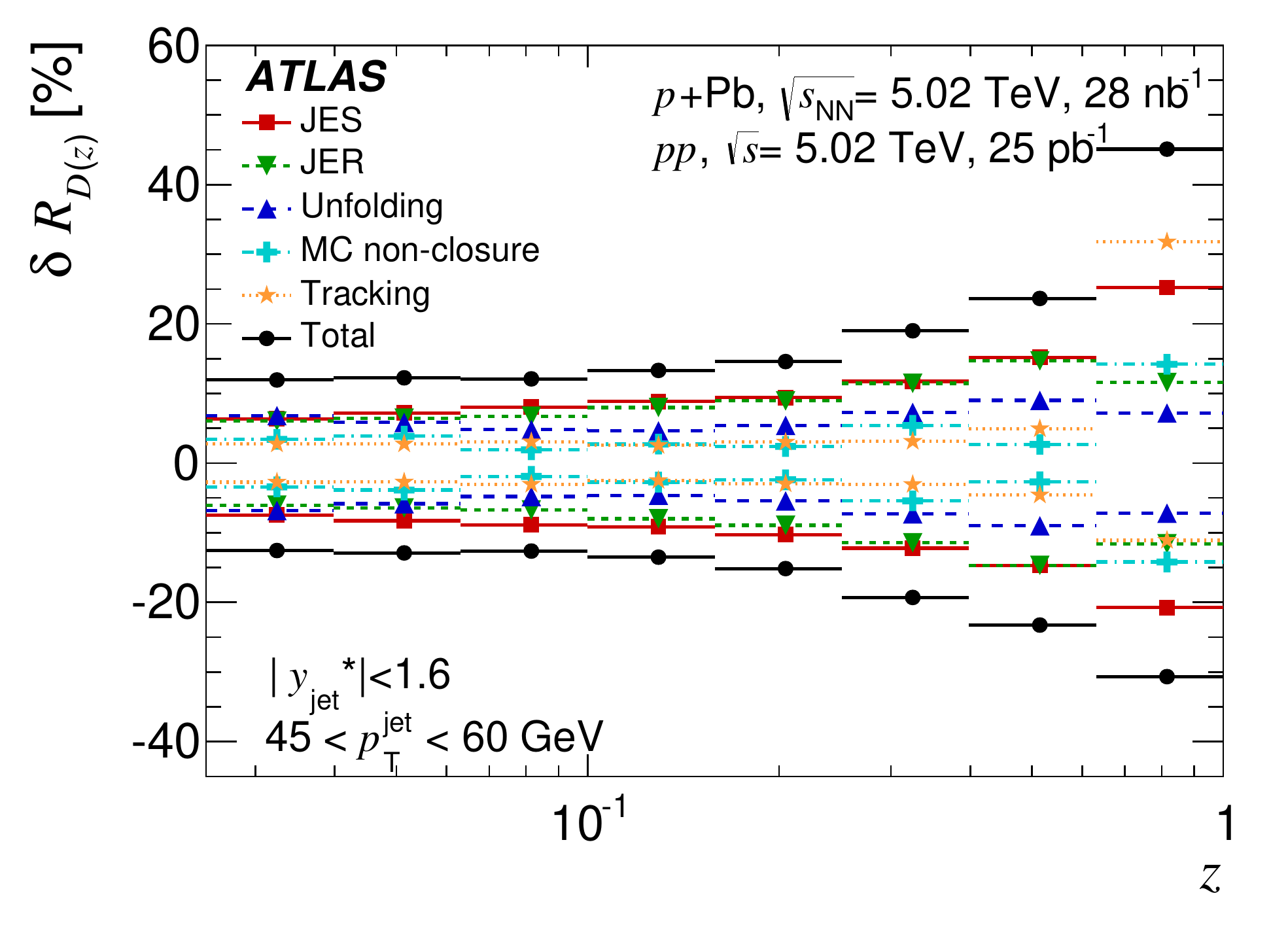} &
\includegraphics[width=0.45\textwidth]{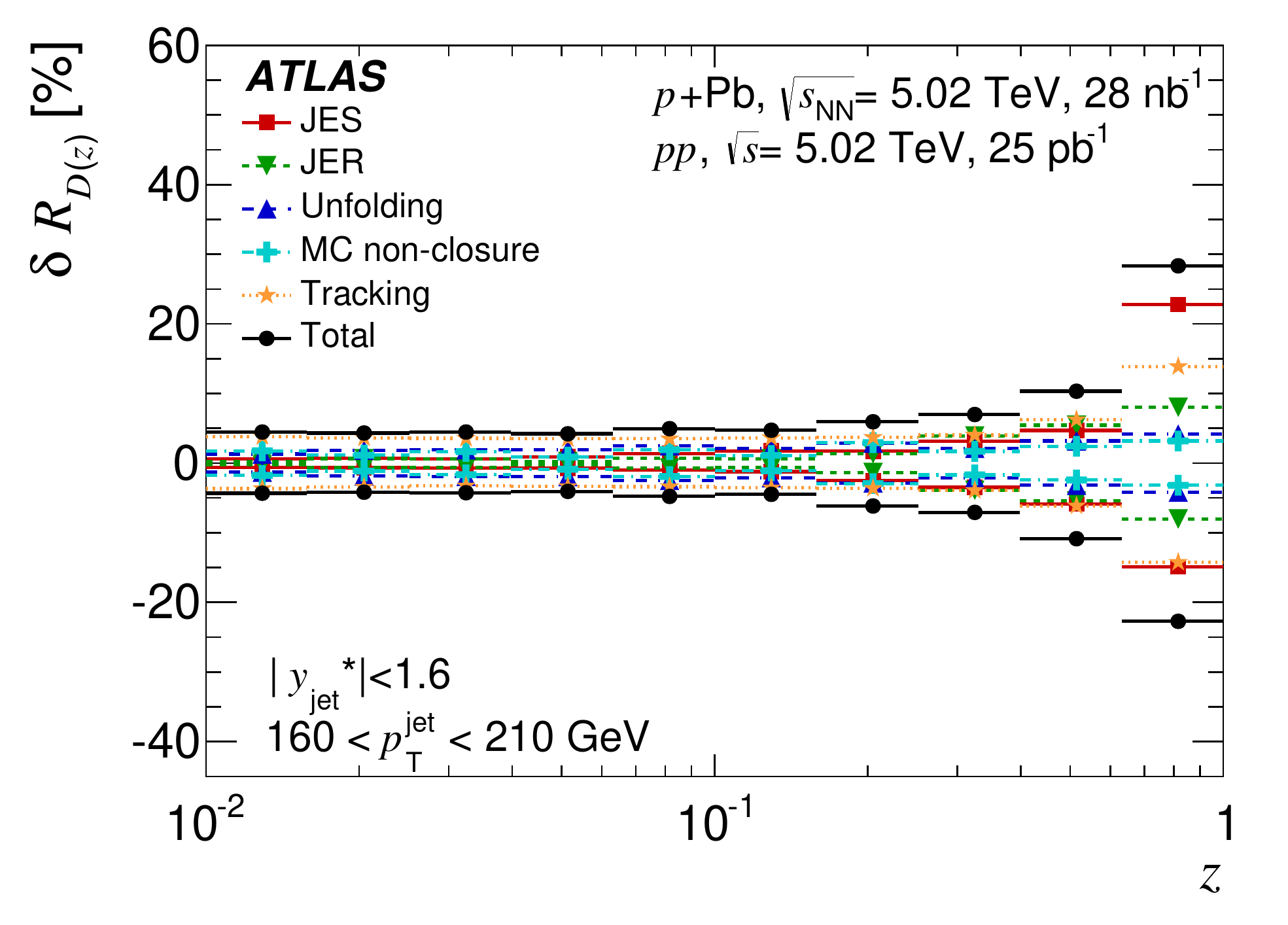} \\
\end{tabular}}
\caption{
Summary of the systematic uncertainties for \Rdz\ ratios, for jets in the 45--60~\GeV\ \pt\ interval (left) and in the 160--210~\GeV\ \pt\ interval (right). The systematic uncertainties due to JES, JER, unfolding, MC non-closure and tracking are shown along with the total systematic
uncertainty from all sources.   
}
\label{Fig:Systematics_3}
\end{figure}

\begin{figure}
\centerline{
\begin{tabular}{cc}
\includegraphics[width=0.45\textwidth]{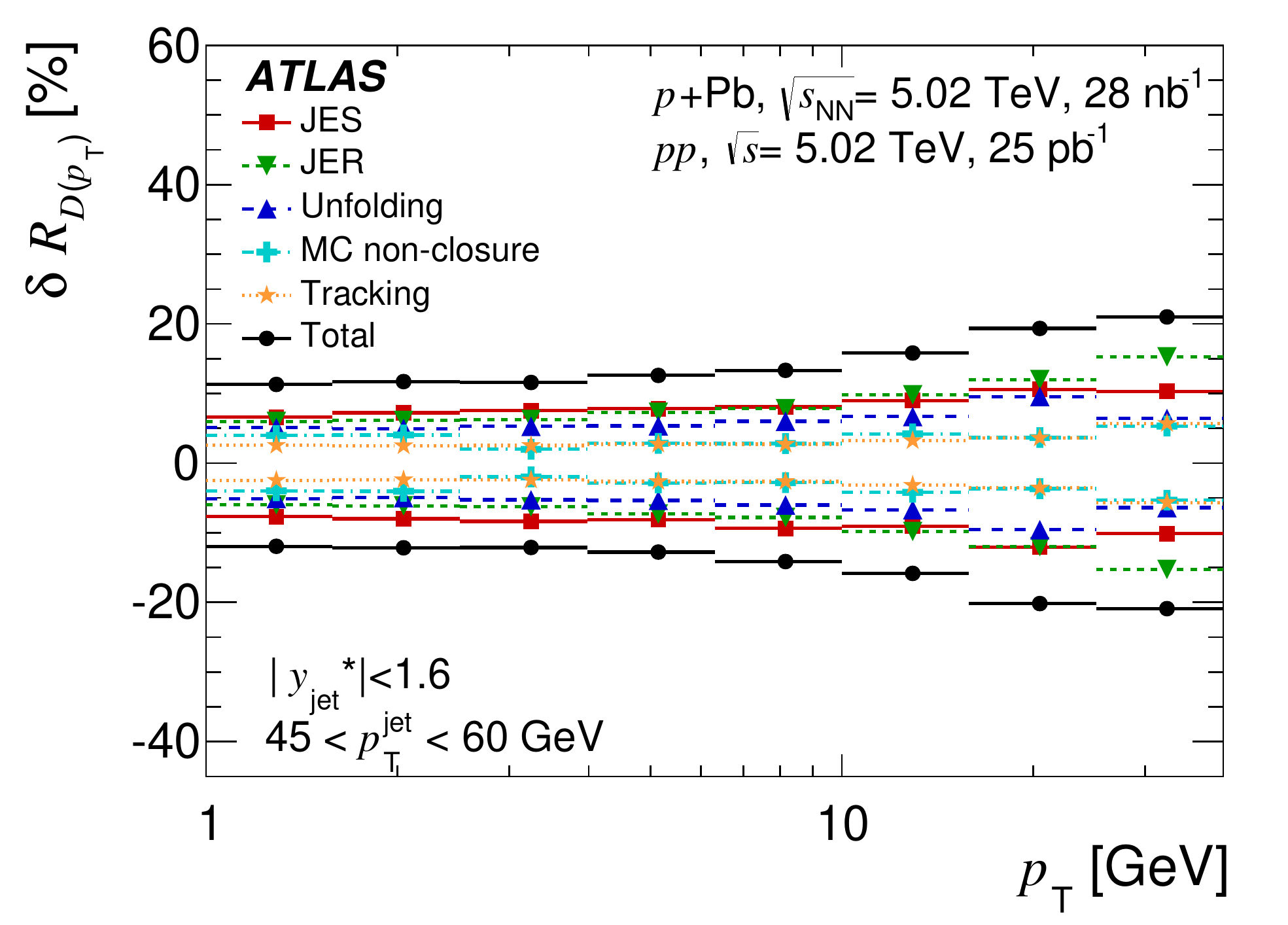} &
\includegraphics[width=0.45\textwidth]{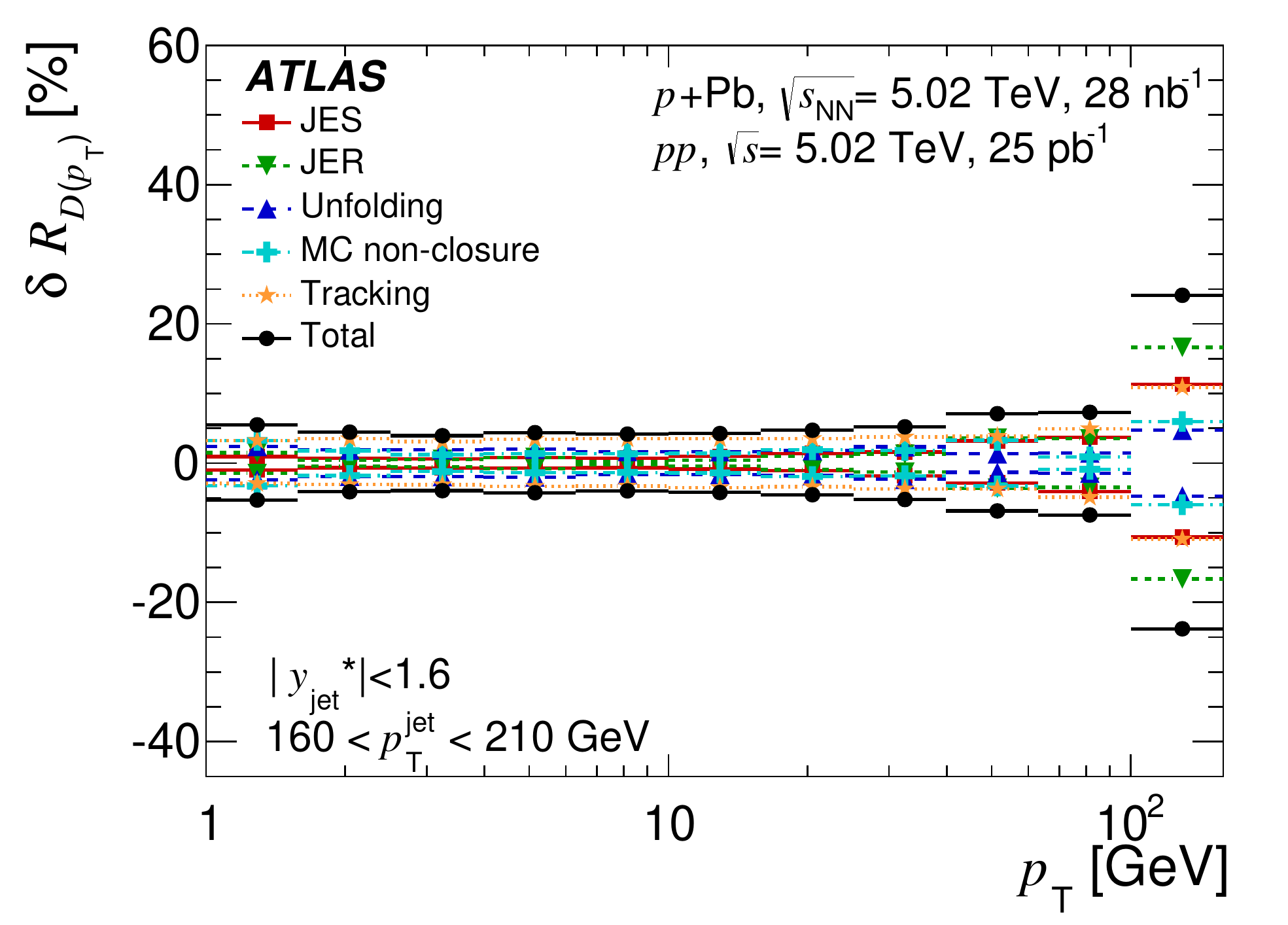} \\
\end{tabular}}
\caption{
Summary of the systematic uncertainties for \Rdpt\ ratios, for jets in the 45--60~\GeV\ \pt\ interval (left) and in the 160--210~\GeV\ \pt\ interval (right). The systematic uncertainties due to JES, JER, unfolding, MC non-closure and tracking are shown
along with the total systematic uncertainty from all sources.
}
\label{Fig:Systematics_4}
\end{figure}

The correlations between the various systematic components in the two collision systems
are considered when taking the ratios of \pPb\ to \pp\ fragmentation functions.
For the JES uncertainty, each source of uncertainty is classified as either correlated or uncorrelated between the two systems depending on its origin.
The JER, unfolding and MC non-closure uncertainties are taken to be uncorrelated.  For the 
tracking-related uncertainties 
the variation in the selection requirements, tracking in dense environments, secondary-track and fake-track rates, and parameterization of the efficiency corrections are taken as 
uncorrelated.  The first three of these are conservatively considered as uncorrelated because the tracking system 
was augmented with the
IBL and the tracking algorithm changed
between the \pPb\ and \pp\ data-taking periods. The uncertainties due to the track-to-particle
matching and the inactive material in the MC samples are taken as correlated between \pPb\ and \pp\ collisions.  
For the correlated uncertainties the ratios are re-evaluated applying the variation to both collision
systems; the resulting variations of the ratios from their central values is used as the correlated systematic uncertainty. The total systematic uncertainties in the \Rdz\ and \Rdpt\ distributions are shown in Figures~\ref{Fig:Systematics_3} and~\ref{Fig:Systematics_4}, respectively, for two \ptjet\ intervals.

%%%%%%%%%%%%%%%%%%%%%%%%%%%
\section{Results}
\label{sec:results}
\begin{figure}
\centerline{
\includegraphics[width=1.0\textwidth]{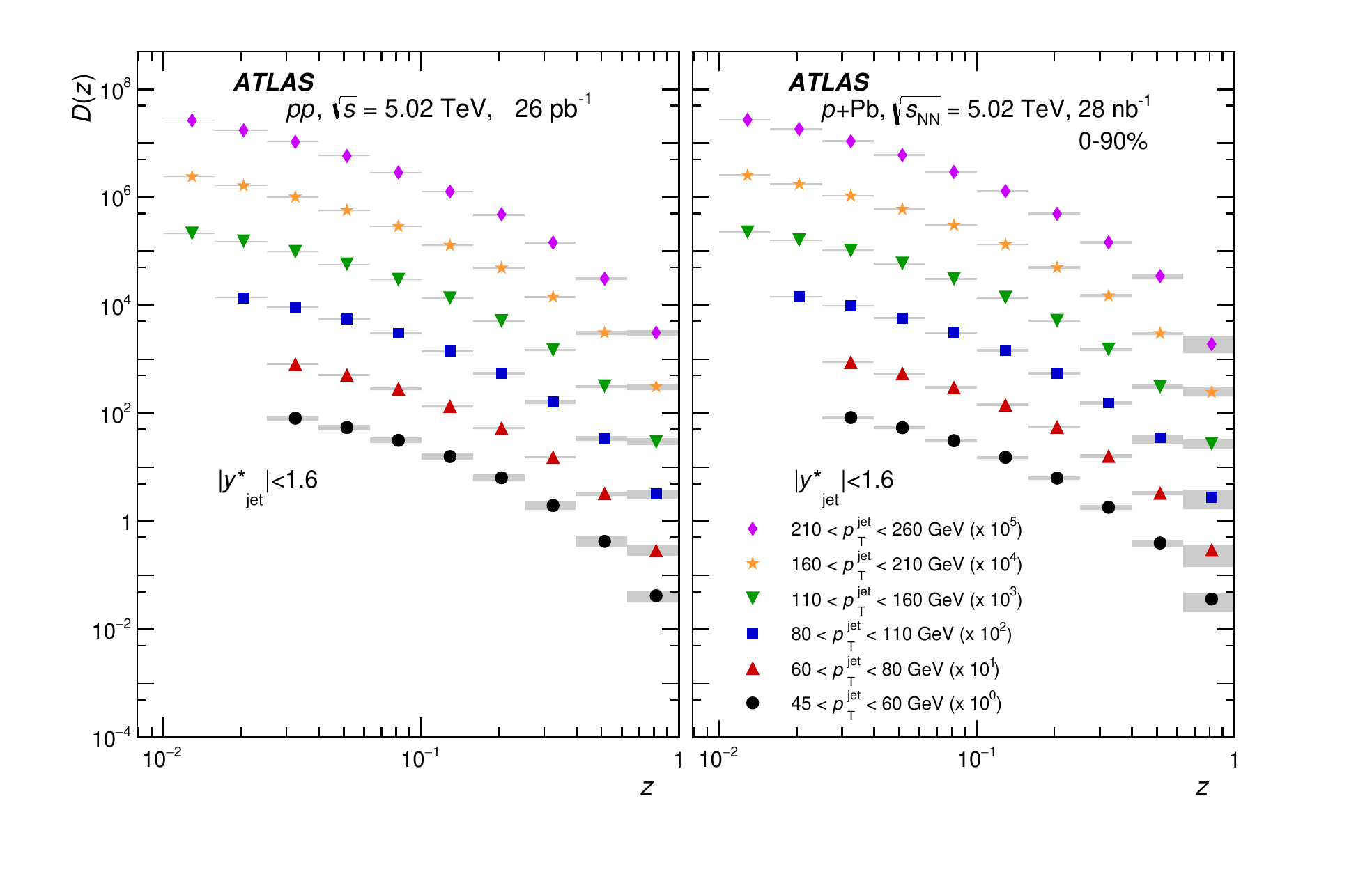}}
\caption{Fragmentation functions as a function of the charged particle \z\ 
   in \pp\ (left) and \pPb\ collisions (right) for the \ptjet\ intervals used in this analysis.
The fragmentation functions in both collision systems are offset by multiplicative factors for clarity as noted
in the legend. 
The statistical uncertainties are shown as error bars and the systematic uncertainties are shown as shaded boxes.  In many
cases the statistical uncertainties are smaller than the marker size.
 }
\label{fig:ff}
\end{figure}

\begin{figure}
\centerline{
\includegraphics[width=1.0\textwidth]{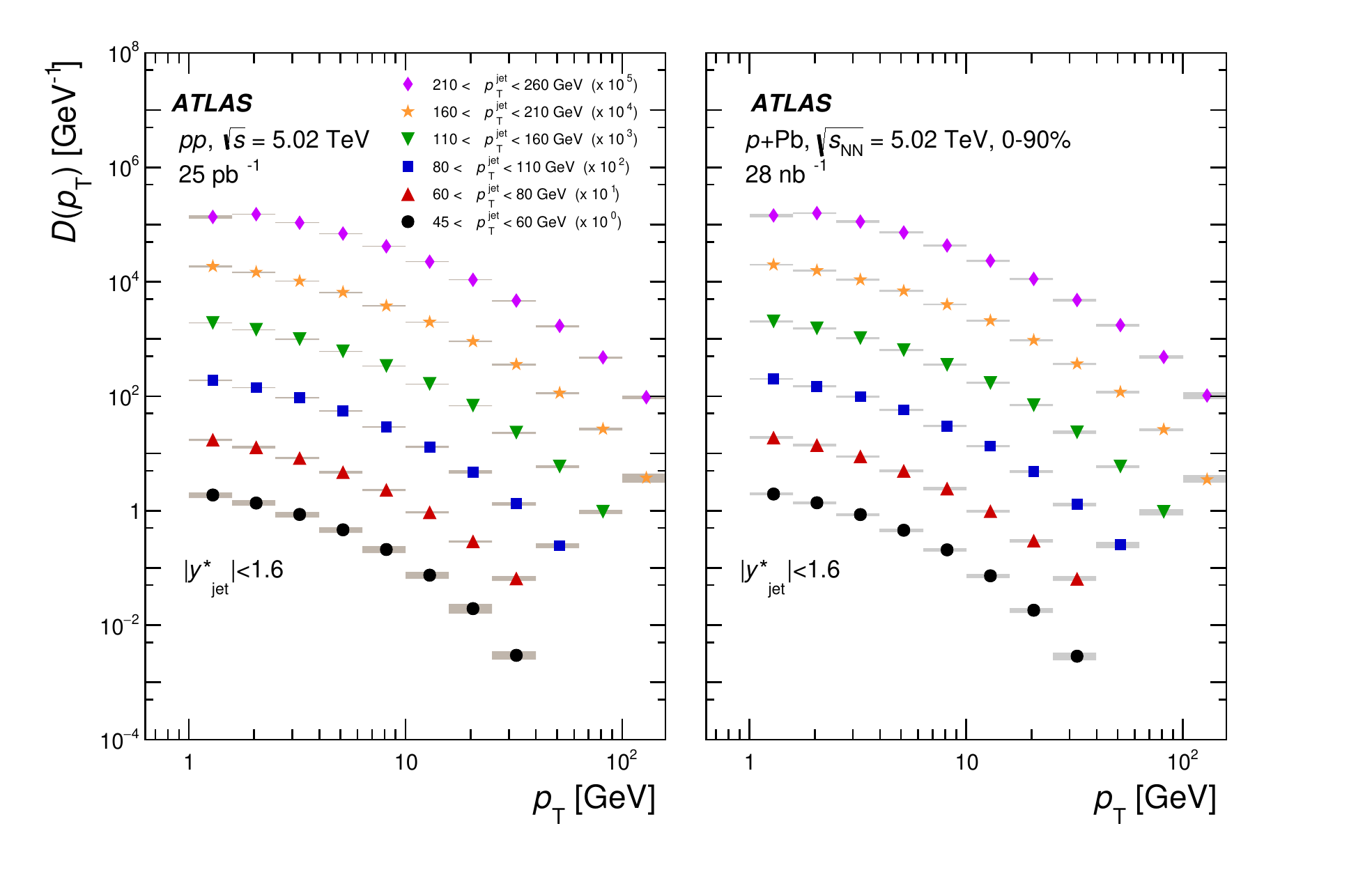}}
\caption{Fragmentation functions as a function of the charged particle \pT\
in \pp\ (left) and \pPb\ collisions (right) for the \ptjet\ intervals used in this analysis.
The fragmentation functions in both collision systems are offset by multiplicative factors for clarity as noted
in the legend. 
The statistical uncertainties are shown as error bars and the systematic uncertainties are shown as shaded boxes.  In many
cases the statistical uncertainties are smaller than the marker size.
 }
\label{fig:ffpt}
\end{figure}

\begin{figure}
\centering{
\includegraphics[width=1.0\textwidth]{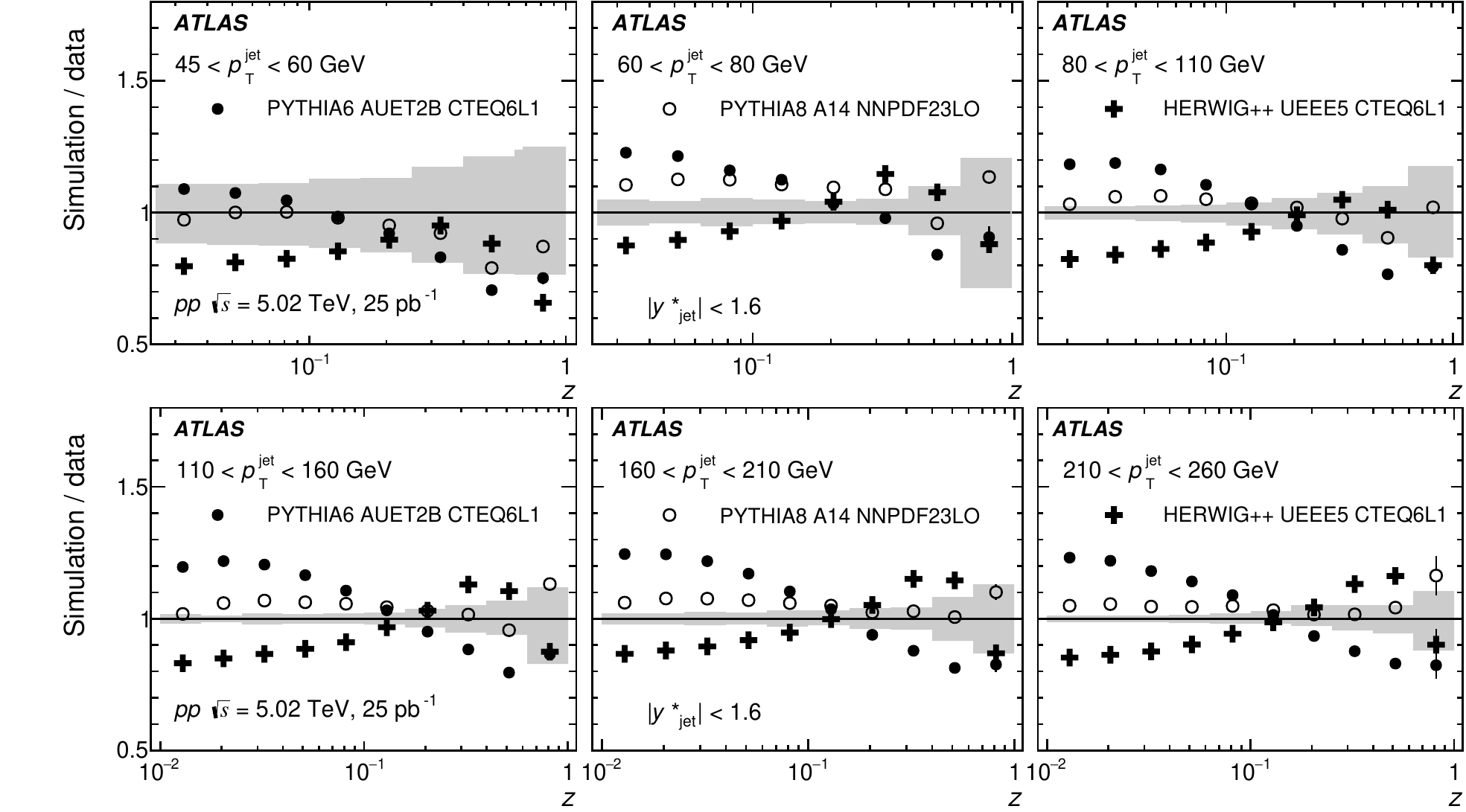}}
\caption{Ratios of the particle-level \Dz\ distributions
   from \pythiasix, \pythiaeight, and \herwig\ to the unfolded \pp\ data
for the six \ptjet\ intervals
used in this analysis. The statistical uncertainties are shown as error bars and 
the systematic uncertainties in the data are shown as the shaded region around unity. In many cases the statistical uncertainties are smaller than the marker
size.}
\label{fig:pp_comp}
\end{figure}

The \Dz\ and \Dpt\ distributions in both collision 
systems are shown in Figures~\ref{fig:ff} and~\ref{fig:ffpt}, respectively.
Figure~\ref{fig:pp_comp} compares the \Dz\ distribution in \pp\ collisions
at 5.02~\TeV\ to the predictions from three event generators (\pythiasix, \pythiaeight, and \herwig)
using the parameter-value tunes and PDF sets described in Section~\ref{sec:data} 
 for the six \ptjet\ intervals.
The \pythiaeight\ generator provides the best description of the data, 
generally agreeing within about 5 to 10\% 
over the kinematic range used here.  \pythiasix\ agrees
within approximately 25\% when 
compared to the data and \herwig\ agrees within approximately 20\%
except for the highest \z\ region, where there are some larger deviations.
Similar agreement with \pythiasix\ and \herwig\ generators with different tunes than used in this analysis was reported by ATLAS in the measurement of fragmentation functions in 7~\TeV\ \pp\ collisions~\cite{Aad:2011sc}. The tunes of \pythiasix\ and \pythiaeight\ used here include the results from that measurement.

\begin{figure}
\centering{
\includegraphics[width=1.0\textwidth]{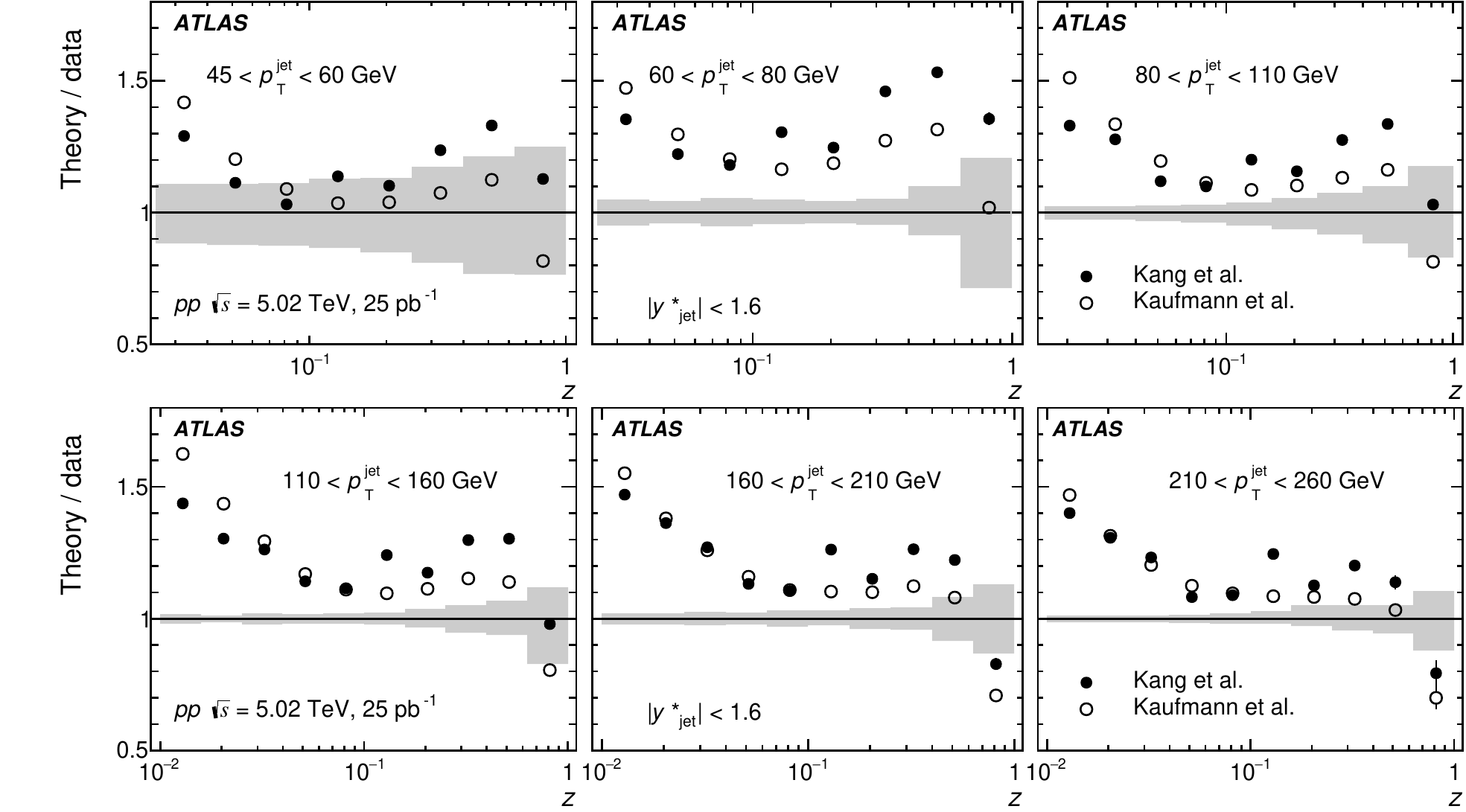}}
\caption{Ratios of theoretical calculations from Refs.~\cite{Kang:2016ehg,Chien:2015ctp} (solid points)
   and Ref.~\cite{Kaufmann:2015hma} (open points) to the unfolded \pp\  \Dz\ distributions
for the six \ptjet\ intervals
used in this analysis. The statistical uncertainties are shown as error bars and 
the systematic uncertainties in the data are shown as the shaded region around unity.
The uncertainties in the theoretical calculations are not shown.}
\label{fig:theory_comp}
\end{figure}

Figure~\ref{fig:theory_comp} shows the \pp\ fragmentation functions compared to two theoretical calculations. These predictions use a slightly different definition of \z\ compared to the definition used in this measurement. This can introduce a~difference between the fragmentation functions of approximately 1\%.    
The calculation in Refs.~\cite{Kang:2016ehg,Chien:2015ctp}
provides fragmentation functions with 
next-to-leading-order (NLO) accuracy as well as a resummation of logarithms in the jet cone size.
The calculation in Ref.~\cite{Kaufmann:2015hma} is at NLO and uses the approximation that the
jet cone is narrow.  
For the parton-to-charged-hadron fragmentation functions, both calculations
use DSS07 fragmentation functions~\cite{deFlorian:2007ekg}.
The uncertainties in the theoretical calculation are not estimated, including the 
uncertainty in DSS07, which is common to both calculations.
The calculations
are systematically higher 
than the data and generally agree within \mbox{20--30\%}.
Larger deviations are observed  at the low and high $z$ regions. 
The DSS07 fragmentation functions have a minimum
\z\ of 0.05 and the calculations use extrapolated fragmentation functions in the region below $z=0.05$.

\begin{figure}
\centerline{
\includegraphics[width=1.0\textwidth]{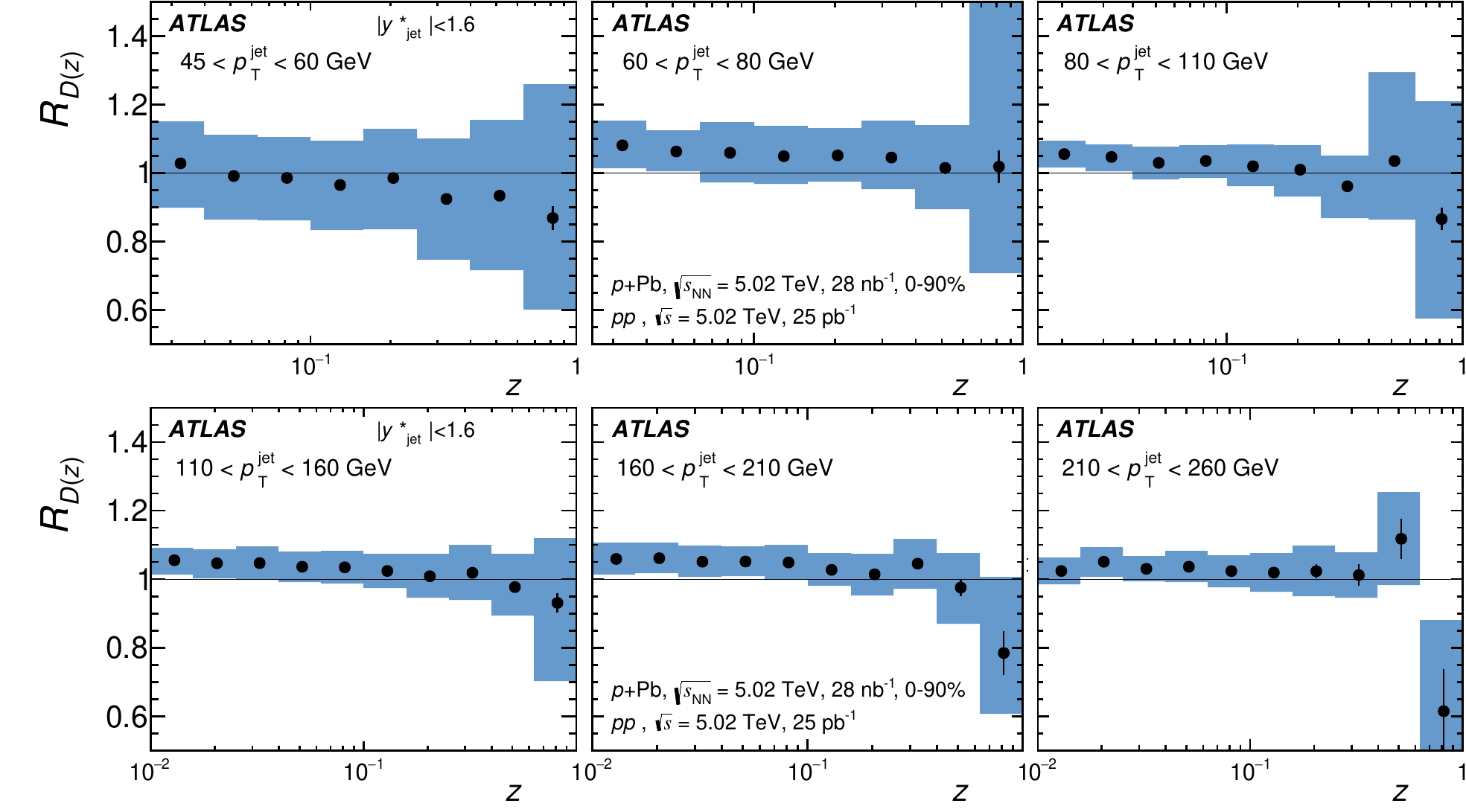} }
\caption{Ratios of fragmentation functions as a function of the charged particle \z\ 
   in \pPb\ collisions to those in \pp\ collisions for
the six \ptjet\ intervals.  The statistical uncertainties are shown as error bars and the total systematic uncertainties are shown as shaded boxes. }
\label{fig:ratios}
\end{figure}

\begin{figure}
\centerline{
\includegraphics[width=1.0\textwidth]{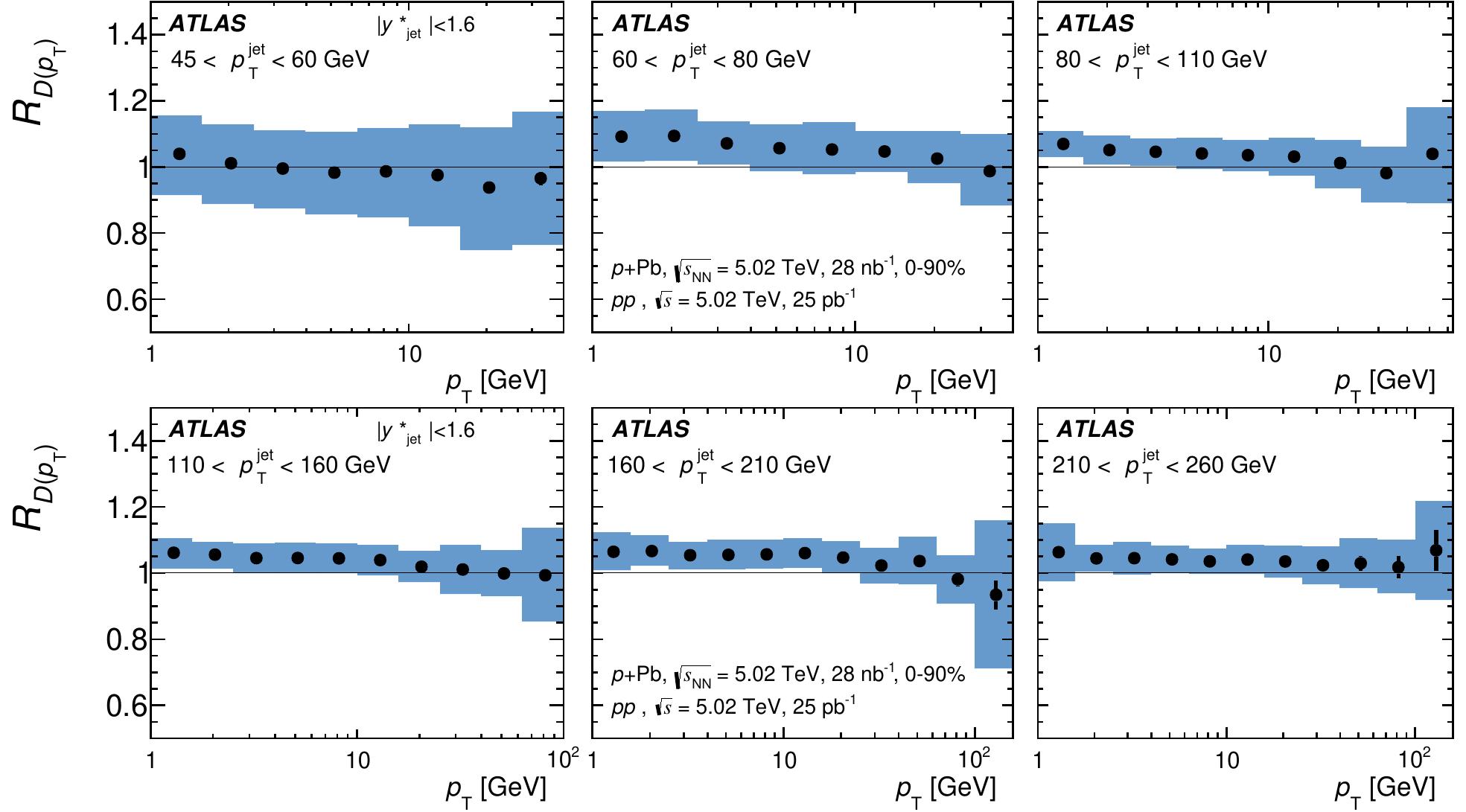} }
\caption{Ratios of fragmentation functions as a function of the charged particle \pT\
   in \pPb\ collisions to those in \pp\ collisions for
the six \ptjet\ intervals.  The statistical uncertainties are shown as error bars and the total systematic uncertainties are shown as shaded boxes. }
\label{fig:ratiospt}
\end{figure}

Figures~\ref{fig:ratios} and \ref{fig:ratiospt} show the ratios of
fragmentation functions in \pPb\ collisions to those in \pp\ collisions, as a function of
\z\ and \pT\ respectively for \ptjet\ from 45 to 260~\GeV.  
Over the kinematic range selected here, the \Rdz\ and \Rdpt\ distributions show deviations from unity of up to approximately 5\% (up to 10\% for \mbox{60--80~\GeV} jet selections)
for \z\ $<$~0.1 and \pT $<$ 10~\GeV.  The deviations are larger than the reported systematic uncertainties by at most a couple of percent and always less than $1.5 \sigma$ of the systematic uncertainties.  
At higher \z\ and \pT\ values the ratios are consistent with unity. At the highest \z\ points for
the \mbox{160--210~\GeV} and \mbox{210--260~\GeV} 
jet selections, deviations from unity of approximately $0.9 \sigma$ and $1.3 \sigma$ of combined statistical and systematic uncertainties, respectively, are observed.
This is not observed in the \Dpt\ distributions.  
In most \ptjet\ bins there is a slight decrease of the central values of \Rdz\ and \Rdpt\ with
increasing \z\ and \pT;  however the size of the effect is smaller than the systematic 
uncertainties.

\FloatBarrier

%%%%%%%%%%%%%%%%%%%%%%%%%%%
\section{Summary}
\label{sec:summary}
This paper presents the first measurement of the jet charged-particle fragmentation functions in a \pA\ collisions system. The jet charged-particle fragmentation functions are reported for $|\ystar|<$~1.6 and \ptjet\ from 45 to 260~\GeV\ in \sqrtsnn~=~5.02~\TeV\ \pPb\ and \pp\ collisions with the \mbox{ATLAS} detector at the LHC. 
The measurement utilizes 28~nb$^{-1}$ of \pPb\ data and 26~pb$^{-1}$ of \pp\ data.
The \pp\ fragmentation functions are compared to predictions from the \pythiasix, \pythiaeight\ 
and \herwig\  generators.  
The generators show deviations from the \pp\ data of up to approximately 25\%, depending on \z\ and the 
choice of generator. \pythiaeight\ with the A14 tune and NNPDF23LO PDF set matches the data most closely.
The \pp\ \Dz\ distributions are also compared to two theoretical calculations based on 
next-to-leading-order QCD and DSS07 fragmentation
functions.  The calculations
are systematically higher 
than the data and agree generally within 20--30\%, with larger deviations at small and large values of \z. 
These measurements help constrain jet fragmentation in \pp\ collisions.
The ratios of fragmentation functions in \pPb\ collisions to those in \pp\ collisions 
show no evidence for modification of jet fragmentation in \pPb\ collisions. 
This measurement provides new constraints on the modifications to jets in \pPb\ collisions at the LHC and
is directly relevant to the current investigations into the properties of small collision systems. Finally, these measurements of jet fragmentation functions for different intervals of jet transverse momentum provide necessary baseline measurements for quantifying the effects of the quark-gluon plasma in \PbPb\ collisions.

\clearpage
%%%%%%%%%%%%%%%%%%%%%%%%%%%

%-------------------------------------------------------------------------------
\section*{Acknowledgments}
%-------------------------------------------------------------------------------

% Acknowledgements for papers with collision data
% Version 14-Feb-2018

% Standard acknowledgements start here
%----------------------------------------------
We thank CERN for the very successful operation of the LHC, as well as the
support staff from our institutions without whom ATLAS could not be
operated efficiently.

We acknowledge the support of ANPCyT, Argentina; YerPhI, Armenia; ARC, Australia; BMWFW and FWF, Austria; ANAS, Azerbaijan; SSTC, Belarus; CNPq and FAPESP, Brazil; NSERC, NRC and CFI, Canada; CERN; CONICYT, Chile; CAS, MOST and NSFC, China; COLCIENCIAS, Colombia; MSMT CR, MPO CR and VSC CR, Czech Republic; DNRF and DNSRC, Denmark; IN2P3-CNRS, CEA-DRF/IRFU, France; SRNSFG, Georgia; BMBF, HGF, and MPG, Germany; GSRT, Greece; RGC, Hong Kong SAR, China; ISF, I-CORE and Benoziyo Center, Israel; INFN, Italy; MEXT and JSPS, Japan; CNRST, Morocco; NWO, Netherlands; RCN, Norway; MNiSW and NCN, Poland; FCT, Portugal; MNE/IFA, Romania; MES of Russia and NRC KI, Russian Federation; JINR; MESTD, Serbia; MSSR, Slovakia; ARRS and MIZ\v{S}, Slovenia; DST/NRF, South Africa; MINECO, Spain; SRC and Wallenberg Foundation, Sweden; SERI, SNSF and Cantons of Bern and Geneva, Switzerland; MOST, Taiwan; TAEK, Turkey; STFC, United Kingdom; DOE and NSF, United States of America. In addition, individual groups and members have received support from BCKDF, the Canada Council, CANARIE, CRC, Compute Canada, FQRNT, and the Ontario Innovation Trust, Canada; EPLANET, ERC, ERDF, FP7, Horizon 2020 and Marie Sk{\l}odowska-Curie Actions, European Union; Investissements d'Avenir Labex and Idex, ANR, R{\'e}gion Auvergne and Fondation Partager le Savoir, France; DFG and AvH Foundation, Germany; Herakleitos, Thales and Aristeia programmes co-financed by EU-ESF and the Greek NSRF; BSF, GIF and Minerva, Israel; BRF, Norway; CERCA Programme Generalitat de Catalunya, Generalitat Valenciana, Spain; the Royal Society and Leverhulme Trust, United Kingdom.

The crucial computing support from all WLCG partners is acknowledged gratefully, in particular from CERN, the ATLAS Tier-1 facilities at TRIUMF (Canada), NDGF (Denmark, Norway, Sweden), CC-IN2P3 (France), KIT/GridKA (Germany), INFN-CNAF (Italy), NL-T1 (Netherlands), PIC (Spain), ASGC (Taiwan), RAL (UK) and BNL (USA), the Tier-2 facilities worldwide and large non-WLCG resource providers. Major contributors of computing resources are listed in Ref.~\cite{ATL-GEN-PUB-2016-002}.
%----------------------------------------------

%\input{acknowledgements/Acknowledgements}

%The \texttt{atlaslatex} package contains the acknowledgements that were valid 
%at the time of the release you are using.
%These can be found in the \texttt{acknowledgements} subdirectory.
%When your ATLAS paper or PUB/CONF note is ready to be published,
%download the latest set of acknowledgements from:\\
%\url{https://twiki.cern.ch/twiki/bin/view/AtlasProtected/PubComAcknowledgements}

%The supporting notes for the analysis should also contain a list of contributors.
%This information should usually be included in \texttt{mydocument-metadata.tex}.
%The list should be printed either here or before the table of contents.

%-------------------------------------------------------------------------------
%\clearpage
%\appendix
%\part*{Appendix}
%\addcontentsline{toc}{part}{Appendix}
%-------------------------------------------------------------------------------

%In a paper, an appendix is used for technical details that would otherwise disturb the flow of the paper.
%Such an appendix should be printed before the Bibliography.

%-------------------------------------------------------------------------------
% If you use biblatex and either biber or bibtex to process the bibliography
% just say \printbibliography here
\printbibliography
% If you want to use the traditional BibTeX you need to use the syntax below.
%\bibliographystyle{bibtex/bst/atlasBibStyleWoTitle}
%\bibliography{mydocument,bibtex/bib/ATLAS}
%-------------------------------------------------------------------------------

%-------------------------------------------------------------------------------
% Print the list of contributors to the analysis
% The argument gives the fraction of the text width used for the names
%-------------------------------------------------------------------------------
\clearpage
%\PrintAtlasContribute{0.30}

%-------------------------------------------------------------------------------
%\clearpage
%\appendix
%\section*{Auxiliary material}
%\addcontentsline{toc}{part}{Auxiliary material}
%\input{auxmaterial.tex}
%-------------------------------------------------------------------------------

%In an ATLAS paper, auxiliary plots and tables that are supposed to be made public 
%should be collected in an appendix that has the title \enquote{Auxiliary material}.
%This appendix should be printed after the Bibliography.
%At the end of the paper approval procedure, this information can be split into a separate document
%-- see \texttt{atlas-auxmat.tex}.

%In an ATLAS note, use the appendices to include all the technical details of your work
%that are relevant for the ATLAS Collaboration only (e.g.\ dataset details, software release used).
%This information should be printed after the Bibliography.

\newpage % ATLAS Collaboration author list
% Reference date of HION-2015-04 is 2016-12-23
% Author list last updated on date 04-SEP-18
% Data extracted on 04-Sep-2018 for paper reference HION-2015-04
% at 9:40am
 
\begin{flushleft}
{\Large The ATLAS Collaboration}

\bigskip

M.~Aaboud$^\textrm{\scriptsize 34d}$,    
G.~Aad$^\textrm{\scriptsize 99}$,    
B.~Abbott$^\textrm{\scriptsize 124}$,    
J.~Abdallah$^\textrm{\scriptsize 75}$,    
O.~Abdinov$^\textrm{\scriptsize 13,*}$,    
B.~Abeloos$^\textrm{\scriptsize 128}$,    
S.H.~Abidi$^\textrm{\scriptsize 164}$,    
O.S.~AbouZeid$^\textrm{\scriptsize 143}$,    
N.L.~Abraham$^\textrm{\scriptsize 153}$,    
H.~Abramowicz$^\textrm{\scriptsize 158}$
% ...
\bigskip
\\

% ...
$^{13}$Institute of Physics, Azerbaijan Academy of Sciences, Baku; Azerbaijan.\\
% ...
$^{34}$$^{(a)}$Facult\'e des Sciences Ain Chock, R\'eseau Universitaire de Physique des Hautes Energies - Universit\'e Hassan II, Casablanca;$^{(b)}$Centre National de l'Energie des Sciences Techniques Nucleaires (CNESTEN), Rabat;$^{(c)}$Facult\'e des Sciences Semlalia, Universit\'e Cadi Ayyad, LPHEA-Marrakech;$^{(d)}$Facult\'e des Sciences, Universit\'e Mohamed Premier and LPTPM, Oujda;$^{(e)}$Facult\'e des sciences, Universit\'e Mohammed V, Rabat; Morocco.\\
% ...
$^{75}$University of Iowa, Iowa City IA; United States of America.\\
% ...
$^{99}$CPPM, Aix-Marseille Universit\'e, CNRS/IN2P3, Marseille; France.\\
% ...
$^{124}$Homer L. Dodge Department of Physics and Astronomy, University of Oklahoma, Norman OK; United States of America.\\
% ...
$^{128}$LAL, Universit\'e Paris-Sud, CNRS/IN2P3, Universit\'e Paris-Saclay, Orsay; France.\\
% ...
$^{143}$Santa Cruz Institute for Particle Physics, University of California Santa Cruz, Santa Cruz CA; United States of America.\\
% ...
$^{153}$Department of Physics and Astronomy, University of Sussex, Brighton; United Kingdom.\\
% ...
$^{158}$Raymond and Beverly Sackler School of Physics and Astronomy, Tel Aviv University, Tel Aviv; Israel.\\
% ...
$^{164}$Department of Physics, University of Toronto, Toronto ON; Canada.\\
% ...
$^{a}$ Also at  Department of Physics, University of Malaya, Kuala Lumpur; Malaysia.\\
$^{b}$ Also at Academia Sinica Grid Computing, Institute of Physics, Academia Sinica, Taipei; Taiwan.\\
$^{c}$ Also at Borough of Manhattan Community College, City University of New York, NY; United States of America.\\
$^{d}$ Also at Centre for High Performance Computing, CSIR Campus, Rosebank, Cape Town; South Africa.\\
$^{e}$ Also at CERN, Geneva; Switzerland.\\
% ...
$^{*}$ Deceased

\end{flushleft}

% Created with Glance <Atlas.Glance@cern.ch>

\end{document}